\documentclass[11pt,a4paper,preprint,amsmath,amssymb,tightenlines]{article}
\usepackage{jheppub}
\usepackage[english]{babel}
\usepackage{graphicx}
\usepackage{bm}
\usepackage{subcaption}
\usepackage{xspace}
\usepackage{hyperref}

\allowdisplaybreaks
\newcommand*{\cf}{cf.\ }
\newcommand*{\ie}{i.\,e.\ }
\newcommand*{\eg}{e.\,g.\ }

\newcommand{\bfp}{\ensuremath{\bm{p}}\xspace}

\title{Analytic next-to-leading order calculation of energy-energy correlation in
gluon-initiated Higgs decays}

\author{Ming-xing Luo,}
\author{Vladyslav Shtabovenko,}
\author{Tong-Zhi Yang,}
\author{and Hua Xing Zhu}

\affiliation{Zhejiang  Institute of Modern Physics, Department of Physics, Zhejiang University, \\ Hangzhou 310027, China\vspace{0.2cm}}

\emailAdd{mingxingluo@zju.edu.cn}
\emailAdd{vshtabov@zju.edu.cn}
\emailAdd{yangtz@zju.edu.cn}
\emailAdd{zhuhx@zju.edu.cn}

\abstract{
The energy-energy correlation~(EEC) function in $e^+e^-$
annihilation is currently the only QCD event shape observable for which
we know the full analytic result at the next-to-leading order (NLO).
In this work we calculate the  EEC observable for gluon initiated Higgs decay  analytically at NLO in the Higgs Effective Field Theory (HEFT) framework and provide the full results expressed in terms of classical polylogarithms, including the asymptotic behavior in the collinear and back-to-back limits.   This observable can be, in principle, measured at the future $e^+e^-$ colliders such as CEPC, ILC, FCC-ee or CLIC. It provides an interesting opportunity to simultaneously probe our understanding of the strong and Higgs sectors and  can be used for the determinations of the strong coupling. 
}

\begin{document}
\maketitle
\flushbottom

\section{Introduction}
\label{sec:introduction} 

Many aspirations of the particle physics community for measuring the properties of the recently discovered Higgs-like boson with unprecedented precision and possibly finding new physics beyond the Standard Model
are closely linked to the prospects of having a new high-energy $e^+e^-$ collider in the near future. 
Precision QCD studies constitute an integral part of the CEPC~\cite{CEPCStudyGroup:2018rmc,CEPCStudyGroup:2018ghi}, ILC~\cite{Behnke:2013xla,Baer:2013cma}, FCC-ee~\cite{Gomez-Ceballos:2013zzn} and CLIC~\cite{Aicheler:2012bya,deBlas:2018mhx} physics programs and it is obvious that our understanding of the strong sector could substantially benefit from new experimental data obtained in the clean environment of a lepton collider.

Over the decades, infrared- and collinear-safe event shape variables turned out  to be very useful to confront theoretical calculations in perturbative QCD (pQCD) with the experimental observations. Being functions of the reconstructed 4-momenta of the final state particles, they can be easily extracted from the experimental data. 
For instance, there are six well-known event shape variables that were studied by the Large Electron Positron (LEP) collider experiments ALEPH~\cite{Heister:2003aj}, DELPHI~\cite{Abdallah:2004xe}, L3~\cite{Achard:2004sv} and OPAL~\cite{Abbiendi:2004qz} in great details. These are thrust~\cite{Brandt:1964sa,Farhi:1977sg}, heavy jet mass~\cite{Clavelli:1981yh}, wide and total jet broadening~\cite{Rakow:1981qn, Ellis:1986ig,Catani:1992jc}, $C$ parameter~\cite{Parisi:1978eg, Donoghue:1979vi} and the jet transition variable $Y_{23}$~\cite{Catani:1991hj}.
Although event shape observables are usually discussed in the context of $e^+e^-$ annihilation, it is also possible to define them in other processes, such as electron-proton~\cite{Aktas:2005tz}, proton-antiproton~\cite{Aaltonen:2011et} or proton-proton~\cite{Banfi:2010xy} collisions. 

On the theory side, fixed-order numerical predictions at next-to-next-to-leading order (NNLO) accuracy are available for a wide range of event shape observables~\cite{Gehrmann-DeRidder:2007nzq,GehrmannDeRidder:2007hr,Weinzierl:2009ms,Ridder:2014wza,DelDuca:2016ily}. Regarding the resummation of endpoint divergences, many results at next-to-next-to-leading logarithmic (NNLL) and even NNNLL accuracy can be found in the literature~\cite{deFlorian:2004mp,
Becher:2008cf,Chien:2010kc,Abbate:2010xh,Monni:2011gb,Becher:2012qc,Banfi:2014sua,Hoang:2014wka,Banfi:2016zlc,Tulipant:2017ybb,Moult:2018jzp,Kardos:2018kqj,Banfi:2018mcq,Bell:2018gce,Verbytskyi:2019zhh}. Publicly available codes such as \textsc{Eerad3}~\cite{Ridder:2014wza} or \textsc{NLOJet++}~\cite{Nagy:2001fj,Nagy:2003tz} and \textsc{Event2}~\cite{Catani:1996jh,Catani:1996vz} implement  parton level predictions for event shape variables at NNLO or NLO accuracy respectively,  while the effects of parton shower and hadronization can be simulated with dedicated software tools.

One of the tasks of a new $e^+e^-$ collider will be to measure various event shape variables with even higher precision as compared to what was possible at LEP. However, one should also take into account that the new machine, being a Higgs factory, will have much higher center of mass energy as compared to LEP.
This opens up many exciting possi\-bilities to calculate and measure event shape variables that would simultaneously probe our understanding of the strong and the Higgs sectors~\cite{Gao:2016jcm}. One of such  observables is the Energy-Energy correlation (EEC) function in gluon-initiated Higgs decays, which is the main subject of this paper.

The original definition of the Energy-Energy Correlation goes back to the late 70s of the last century, when the authors of~\cite{Basham:1978bw} suggested to employ two calorimeters separated by an angle $\chi$ to measure the energies of two hadrons $a$ and $b$ produced in $e^+e^-$ annihilation
\begin{equation}
e^+ (k_1) \,  e^- (k_2) \to \gamma^\ast/ Z^0 \to a (p_a) \, b(p_b) + X. \label{eq:eecpartonic}
\end{equation}
The EEC is a differential angular distribution that arises from measuring the cosine of the angle $\theta_{ab}$ between all particle pairs $(a,b)$ in the event and weighting each contribution by the particle energies $E_a$ and $E_b$. The resulting histogram is normalized to unit area and provides a useful way to visualize the energy flow through the calorimeters. The EEC is defined as
\begin{equation}
 \frac{1}{\sigma_{\textrm{tot}}} \frac{d \Sigma  (\chi)}{d \cos\chi} =
\sum_{a,b} \int \, \frac{2 E_a E_b}{Q^2} \, \delta( \cos\theta_{ab} -
  \cos\chi) \,  d \sigma_{a+b+X}, \label{eq:eecdef}
\end{equation}
where $d \sigma_{a+b+X}$ is the differential production cross section for the process in eq.\,\eqref{eq:eecpartonic}, $Q$ is the total center of mass energy and $\cos\theta_{ab} = \hat{\bm{p}}_a \cdot \hat{\bm{p}}_b$. Here $\sigma_{\textrm{tot}}$ denotes the total cross section for $e^+ e^- \to \textrm{hardons}$. The summation is over all $(a,b)$-combinations from the final state hadrons. Notice that we do not include the contributions from self-correlations in our fixed order calculation such that in our results the summation over $a$ and $b$ was replaced by $\sum_{a \neq b}$. The fixed order pQCD predictions for EEC are obtained by calculating eq.\,\eqref{eq:eecdef}
at the parton level, with the LO contribution arising from the tree level hard process
\begin{equation}
e^+ (k_1) \,  e^- (k_2) \to \gamma^\ast/ Z^0 \to q (p_1) \, \bar{q}(p_2) \, g (p_3).
\end{equation}

At a future Higgs factory, it appears very natural to measure the EEC in hadronic decays of the Higgs boson. Higgs can decay to a pair of gluons through a top quark loop, or to $b\bar{b}/c\bar{c}$ directly via Yukawa couplings. In this paper we are mainly concerned with the gluon initiated decay of the Higgs boson within the framework of the Higgs Effective Field Theory (HEFT)~\cite{Wilczek:1977zn,Shifman:1978zn,Inami:1982xt,Kniehl:1995tn}. The $b$ or $c$ quark Yukawa initiated decay will be considered in a future work. In the HEFT, the top quark can be integrated out to obtain operators that contain the Higgs field $H$ and two QCD field-strength tensors $G^{\mu \nu}$. The interacting part of the HEFT Lagrangian is then given by
\begin{equation}
\label{eq:heft}
\mathcal{L}_{\textrm{HEFT}} = - \frac{1}{4} \lambda H \, \textrm{Tr} \left ( G^{\mu \nu}  G_{\mu \nu} \right ).
\end{equation}
The corresponding Wilson coefficient $\lambda$ is determined from matching the amplitude for $H \to gg$ in SM to the one in HEFT order by order in the strong coupling $\alpha_s$ and the inverse of the top quark mass $1/m_t$. To lowest order in $\alpha_s$ and $1/m_t$ we have 
\begin{equation}
\lambda = \frac{\alpha_s}{3\pi } \sqrt{\sqrt{2} G_F},
\end{equation}
with $G_F$ being the Fermi constant. Currently, $\lambda$ is known at the $\textrm{N}^4\textrm{LO}$~\cite{Baikov:2016tgj} accuracy.  The effective operator in eq.\,\eqref{eq:heft} gives rise to tree-level coupling of the Higgs with $2$, $3$, and $4$ gluons. In this work, we would like to consider the correlations between energies of the partons that arise from a gluon-initiated decay of the Higgs particle at rest, via
\begin{equation}
H \to q (p_1) \, \bar{q}(p_2) \, g (p_3)
\end{equation}
and 
\begin{equation}
H \to g (p_1) \, g(p_2) \, g (p_3).
\end{equation}
To avoid possible confusion between the EEC from eq.\,\eqref{eq:eecdef} and the one considered in this paper, in the following we will denote them as ``standard EEC'' and ``Higgs EEC'' respectively. Then, in analogy to eq.\,\eqref{eq:eecdef} we can define the Higgs EEC as
\begin{equation}
 \frac{1}{\Gamma_{\textrm{tot}}} \frac{d \Sigma_H  (\chi)}{d \cos\chi} =
\sum_{a,b} \int \, \frac{2 E_a E_b}{m_H^2} \, \delta( \cos\theta_{ab} -
  \cos\chi) \,  d \Gamma_{a+b+X},
\end{equation}
where $m_H$ is the Higgs boson mass, $\Gamma_{\textrm{tot}}$ is  the total decay width for $H \to gg$ and 
$d \Gamma_{a+b+X}$ denotes the differential decay rate for Higgs decaying into gluons plus anything else. Notice that here
the center-of-mass energy squared in $e^+ e^-$ annihilation $Q^2$ was replaced with $m_H^2$, since we are considering the decay of the Higgs particle at rest, so that $Q = (m_H,0,0,0)$. The normalization of the Higgs EEC with respect to 
\begin{equation}
\Gamma_{\textrm{tot}} = \frac{\lambda^2 m_H^3}{8 \pi} K(\mu)
\label{eq:hdecaytot}
\end{equation}
ensures the cancellation of $\lambda$ in the final result for the Higgs EEC. The factor $K$ accounts for the corrections to the total decay width $H \to g g$ within the HEFT. For our purposes it is sufficient to use its NLO value~\cite{Chetyrkin:1997iv}
\begin{equation}
K(\mu) = 1 + \frac{\alpha_s}{ 2 \pi} \left [ \frac{73}{2} + 11 \log \frac{\mu^2}{m_H^2} - N_f \left ( \frac{7}{3} + \frac{2}{3} \log \frac{\mu^2}{m_H^2} \right ) \right ] + \mathcal{O}(\alpha_s^2),
\label{eq:hdecaykfac}
\end{equation}
where $N_f$ is the number of light quark flavors.

The derivation of fixed-order pQCD predictions for event shape variables is usually done using numerical methods. While for some observables such as EEC or thrust the LO predictions can be obtained analytically by directly calculating the 3-particle phase space integrals, this ``brute-force'' approach is clearly not feasible at NLO and beyond. At NLO the main complication arises from the evaluation of the double real emissions. The corresponding 4-particle phase space integrals are very difficult to calculate analytically due to the complexity of the integrand and its dependence on a (often nontrivial) measurement function. Even though the final result for the given infrared- and collinear-safe event shape variable must be finite, the single integrals will suffer from severe soft and collinear divergences. The cancellation of the divergences occurs only at the very end, when real-virtual and double-real contributions are added together.

The availability of reliable numerical NNLO predictions for event shape observables in $e^+ e^-$ annihilation is undoubtedly useful, but the corresponding analytic results (even at NLO) are still interesting and desirable. Currently, the standard EEC is the only QCD event shape observable known analytically at NLO~\cite{Dixon:2018qgp}. There has also been remarkable progress in computing EEC analytically in ${\cal N}=4$ supersymmetric Yang-Mills theory~\cite{Belitsky:2013xxa,Belitsky:2013bja}. Analytic results are now available not just at NLO~\cite{Belitsky:2013ofa}, but also at NNLO very recently~\cite{Henn:2019gkr}. The goal of this paper is to obtain the analytic NLO result also for the Higgs EEC using the same methods as in~\cite{Dixon:2018qgp}. To the best of our knowledge, no previous numerical or analytic studies of this observable exist in the literature, even at LO. However, we would like to point out that numerical NLO and approximate NNLO results for thrust in hadronic Higgs decays were recently obtained in~\cite{Gao:2019mlt}.

This work is organized in the following way. In section \ref{sec:calculation} we describe our framework used to calculate EEC-like observables such as the standard EEC or the Higgs EEC analytically at LO and NLO. In particular, we explain the nontrivial procedure of deriving IBP equations for nonlinear propagators and fixing the boundary conditions in the differential equations for the master integrals. Our analytic results for the Higgs EEC are presented in section \ref{sec:fullres}, while section \ref{sec:asymptotics} is devoted to the discussion of the asymptotics in the collinear and back-to-back limits. Section \ref{sec:numerics} contains a comparison of the NLO result to the predictions of \textsc{Pythia} as a way to estimate the importance of the nonperturbative corrections for our observable. Additionally, we employ the \textsc{Pythia} simulation as a toy model for the determination of the strong coupling, which should emphasize the importance of the Higgs EEC for the future, when real data from a new lepton collider should become available. Finally, we summarize the obtained results and present an outlook for future work in this direction in section \ref{sec:summary}.

For convenience, in the course of our calculation we set $m_H$ to unity. This is allowed, as the dependence of the final result on the Higgs mass can be easily restored by dimensional analysis. In order to avoid cluttering the notation, we prefer to use $m_H\equiv1$ in all relations shown in section \ref{sec:calculation}. Notice that in this case the scalar products $p_a \cdot Q$ and $p_b \cdot Q$ have mass dimension one.  The formulas provided in section \ref{sec:fullres} and all subsequent sections do not make use of this simplficiation and display the full dependence on $m_H$.

\section{Calculational setup}
\label{sec:calculation} 

\subsection{Outline}
\label{sec:outline} 

To approach the task outlined in section \ref{sec:introduction} we directly calculate the matrix elements squared of all the relevant sub-processes and convert the arising 4-particle phase space integrals into loop integrals using the method of reverse unitarity~\cite{Anastasiou:2002yz,Anastasiou:2003yy}. Those integrals are then reduced to a smaller set of master integrals via Integration-By-Parts reduction~\cite{Chetyrkin:1981qh,Tkachov:1981wb}, while for the evaluation of the master integrals we employ the method of differential equations~\cite{Kotikov:1991pm,Kotikov:1990kg,Kotikov:1991hm,Bern:1993kr,Remiddi:1997ny,Gehrmann:1999as}. Each system of equations is calculated by first turning it to the canonical form~\cite{Henn:2013pwa} and then fixing the boundary constants.

Among all QCD event shape observables, EEC is arguably the simplest object that can be computed in this framework. Unlike thrust, jet masses, jet broadening and the jet transition variable $Y_{23}$, the measurement function of EEC 
\begin{equation}
E_a E_b \, \delta( \cos\theta_{ab} -  \cos\chi) \label{eq:mf}
\end{equation}
does not require us to minimize or maximize a particular kinematic quantity. Furthermore, the measurement function in the contribution from the particle pair $(a,b)$ depends only on the energies and momenta of these particles, which greatly facilitates the calculation of the individual contributions. However, since
\begin{equation}
\cos \theta_{ab} = 1 - \frac{p_a \cdot p_b}{p_a \cdot Q \, p_b \cdot Q},
\end{equation}
applying the reverse unitarity~\cite{Anastasiou:2002yz,Anastasiou:2003yy}
\begin{equation}
\frac{d^d p}{(2 \pi)^d} 2 \pi \delta_+ (p^2 - m^2)\to \frac{1}{i} \frac{d^d p}{(2 \pi)^d} \left ( \frac{1}{p^2 - m^2 - i \varepsilon}  - \frac{1}{p^2 - m^2 + i \varepsilon} \right )
\end{equation}
to eq.\,\eqref{eq:mf} yields a propagator that is not linear in the loop momentum dependent scalar products. At this point it is convenient to introduce a new dimensionless  variable $z$ so that 
\begin{equation}
2 z \equiv 1 - \cos \chi
\end{equation}
and
\begin{equation}
E_a E_b \, \delta( \cos\theta_{ab} -  \cos\chi) = (p_a \cdot Q)^2 (p_b \cdot Q)^2 \delta \left (
2 z \, p_a \cdot Q \, p_b \cdot Q - p_a \cdot p_b \right ).
\end{equation}
This property of the measurement function is the reason why the analytic NLO calculation of an EEC event shape variable is nontrivial. At first glance, the complications related to the appearance of the nonlinear propagator
\begin{equation}
\frac{1}{2 z \, p_a \cdot Q \, p_b \cdot Q - p_a \cdot p_b} \biggl |_{\textrm{cut}} \label{eq:nonlinprop}
\end{equation}
make it very challenging to carry out the calculation using the standard techniques. Yet, as will be explained in the following sections, all these difficulties can be efficiently mitigated with a minimal amount of out of the box thinking.

The main feature of our approach is the direct IBP reduction of loop integrals with nonlinear propagators and the subsequent calculation of the corresponding master integrals. A different route was taken in~\cite{Gituliar:2017umx}, where the nonlinear propagator was converted into a product of two linear propagators at the price of introducing an auxiliary parameter $x$. While this linearization significantly facilitates the IBP reduction of the phase space integrals, the evaluation of the auxiliary master integrals using differential equations seems to become more involved. Nevertheless, it would be very interesting to compare the results in our approach~\cite{Dixon:2018qgp} with those using the method of Ref.~\cite{Gituliar:2017umx}, once they become available. 

\subsection{Amplitudes}

\begin{figure}
  \begin{subfigure}[b]{0.45\textwidth}
    \includegraphics[width=\textwidth]{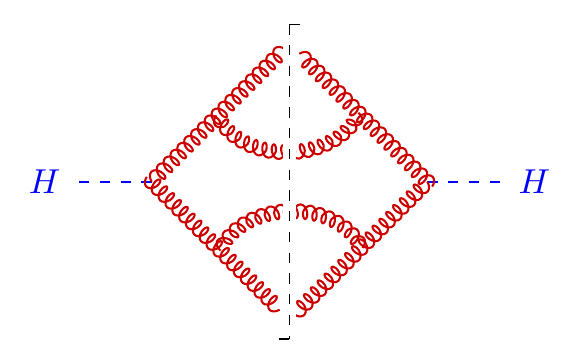}
    \caption{$gggg$}
  \end{subfigure}
    \begin{subfigure}[b]{0.45\textwidth}
    \includegraphics[width=\textwidth]{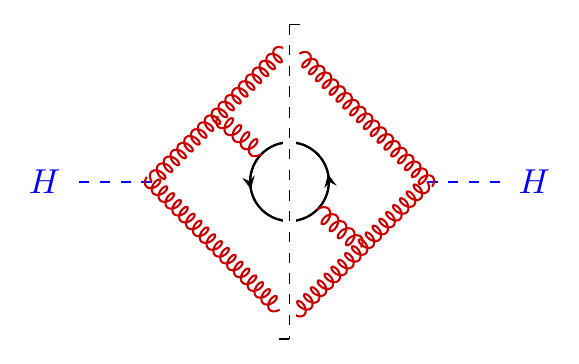}
    \caption{$gg q \bar{q}$}
  \end{subfigure}  
  \begin{subfigure}[b]{0.47\textwidth}
    \includegraphics[width=\textwidth]{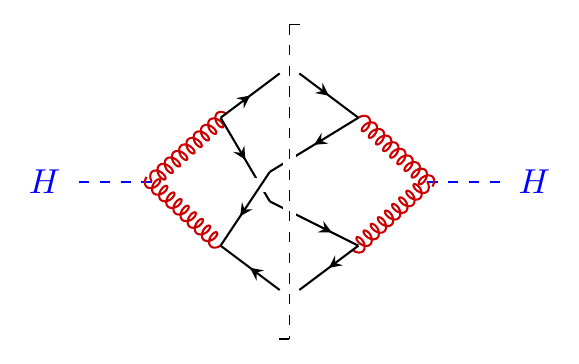}
    \caption{$q \bar{q} q \bar{q}$}
  \end{subfigure}
  \hfill
  \begin{subfigure}[b]{0.47\textwidth}
    \includegraphics[width=\textwidth]{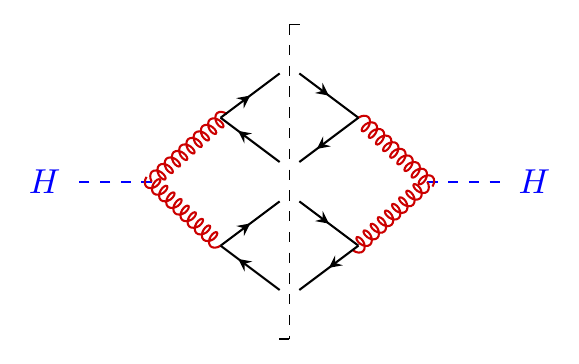}
    \caption{$q \bar{q} q' \bar{q}'$}
  \end{subfigure}
  \caption{Some of the representative cut diagrams for real corrections to the Higgs EEC at NLO. For simplicity we do not show diagrams arising from the Higgs couplings to triple and quartic gluon vertices.}
    \label{fig:higgsdiags}
 \end{figure}

The main ingredient of this calculation are the squared matrix elements $|\mathcal{M}(H \to gg + X)|^2$ for the real emission processes 
\begin{subequations}
\begin{align}
H(Q) &\to g(p_1) \, g(p_2) \, g(p_3), \\
H(Q) &\to q(p_1) \, \bar{q}(p_2) \, g(p_3)
\end{align}
\end{subequations}
with 3-parton final states and
\begin{subequations}
\begin{align}
H(Q) &\to g(p_1) \, g(p_2) \, g(p_3) \, g(p_4), \\
H(Q) &\to q(p_1) \, \bar{q}(p_2) \, g(p_3) \, g(p_4), \\
H(Q) &\to q(p_1) \, \bar{q}(p_2) \, q(p_3) \, \bar{q}(p_4), \\
H(Q) &\to q(p_1) \, \bar{q}(p_2) \, q'(p_3) \, \bar{q}'(p_4)
\end{align}
\end{subequations}
with 4-parton final states. To ensure the correctness of the expressions, we generate the corresponding amplitudes in two different ways, using \textsc{QGRAF}~\cite{Nogueira:1991ex} and \textsc{FeynArts}~\cite{Hahn:2000kx}, \cf figure \ref{fig:higgsdiags}.

The \textsc{FeynArts} model for HEFT was created with \textsc{FeynRules}~\cite{Alloul:2013bka} and the Feynman rules for the Higgs-gluon vertices were additionally rederived using the \texttt{FeynRule}  function of  \textsc{FeynCalc}~\cite{Mertig:1990an,Shtabovenko:2016sxi}. The same vertices are also employed when obtaining amplitudes from the output of \textsc{QGRAF}.

Custom \textsc{FORM}~\cite{Vermaseren:2000nd} code is used to evaluate and simplify the squared matrix elements in $d$-dimensions, where the color algebra is handled by the \textsc{Color}~\cite{vanRitbergen:1998pn} package. Summations over the polarizations of the gluons are done using the axial gauge
\begin{equation}
\sum_{\lambda=1}^2 \varepsilon^\mu (\bfp_i, \lambda) \varepsilon^{\ast \nu} (\bfp_i, \lambda) =  -g^{\mu \nu} + \frac{(p_i^\mu n^\nu + p_i^\nu n^\mu)}{p_i \cdot n} - \frac{n^2 p_i^\mu p_i^\nu}{(p_i \cdot n)^2},
\end{equation}
where for a particular gluon with the 4-momentum $p_i$ we use the 4-momentum of another gluon $p_j$ as the auxiliary vector $n$.

Some of the terms in $|\mathcal{M}(H \to gg + X)|^2$ contain linearly dependent propagators that require partial fractioning.  This is also handled by our \textsc{FORM} code, using rules obtained with the \texttt{ApartFF} function of \textsc{FeynCalc}. \texttt{ApartFF}  is based on the multiloop partial fractioning algorithm from~\cite{Feng:2012iq}, which is also implemented in the standalone \textsc{\$Apart} package.

\subsection{Topology identification}

 As far as the topology identification is concerned, the presence of the nonlinear propaga\-tor eq.\,\eqref{eq:nonlinprop} prevents us from using the established methods~\cite{Pak:2011xt} that rely on the $UF$-representation of the loop integrals. Instead, we identify the identical topologies by trying different combinations of loop momentum renamings $p_a \leftrightarrow p_b$ and shifts $p_a \to Q - \sum_{b \neq a} p_b$. Since the sum of the measurement functions is manifestly invariant under such transformations, the topology identification proceeds by considering the full expression
\begin{equation}
\int \left (\prod_k \frac{d^d p_k}{(2\pi)^{d-1}}  \delta_+(p_k^2) \right) |\mathcal{M}(H \to gg + X)|^2 \sum_{a<b} \, 2 E_a E_b \, \delta (\cos \theta_{ab} - \cos \chi)
\end{equation}
and applying all the allowed shifts and renamings to each distinct denominator and its loop momentum dependent coefficient. This allows us to identify all subtopologies contained in
\begin{equation}
\left ( \prod_k \frac{d^d p_k}{(2\pi)^{d-1}} \delta_+(p_k^2) \right) |\mathcal{M}(H \to gg + X)|^2 \, 2 E_a E_b.
\end{equation}
A subtopology is not the final topology of an integral family, since it lacks the nonlinear propagator. Therefore, each subtopology gives rise to
$\left ( \substack{k \\ 2} \right )$ integral families, where $k$ is the number of the final state partons. For example, at LO the single subtopology yields
$1 \times 3$ integral families.

In the case of the subprocess $H \to g g g g$ we slightly modify this procedure to maximally exploit the symmetry of the corresponding matrix element squared under $p_a \leftrightarrow p_b$. This symmetry allows us to write
\begin{align}
\int & \left(\prod_{k=1}^4 \frac{d^d p_k}{(2\pi)^{d-1}} \delta_+(p_k^2)\right) |\mathcal{M}(H\to g g g g)|^2 \sum_{a<b} \, 2 E_a E_b \,  \delta (\cos \theta_{ab} - \cos \chi) \nonumber  \\
& = 
6 \int \left(\prod_{k=1}^4 \frac{d^d p_k}{(2\pi)^{d-1}} \delta_+(p_k^2)\right) |\mathcal{M}(H\to g g g g)|^2 \, 2 E_1 E_2 \, \delta (\cos \theta_{12} - \cos \chi),
\label{eq:topoidsym}
\end{align}
so that in this case the number of the final topologies  equals the number of the subtopologies. Of course, in this case the set of possible shifts and replacements becomes more restricted. For example, transformations such as $p_1 \leftrightarrow p_3$ or $p_2 \leftrightarrow p_4$ would introduce additional measurements functions in eq.\,\eqref{eq:topoidsym}, which we would like to avoid.
On the other hand, the renamings $p_1 \leftrightarrow p_2$ or $p_3 \leftrightarrow p_4$ do not alter the measurement function in eq.\,\eqref{eq:topoidsym} and are, therefore, allowed. This fully symmetric topology identification gives us only 10 integral families in the subprocess $H \to g g g g$ as compared to 18 families when using the other method. Unfortunately, other subprocesses of the Higgs decay do not possess such a high degree of symmetry on the level of the matrix element squared so that the fully symmetric topology identification is not helpful there.

All the manipulations related to the topology identification are handled by an in-house  \textsc{Mathematica} code. Owing to the small number of the loop momenta (at most 4) and the fact that all quarks are taken massless, on a modern laptop this procedure requires only few minutes per subprocess. For the sake of completeness, all identified subtopologies at LO and NLO are listed in appendix \ref{sec:appendix0}.

\subsection{IBP reduction}
\label{sec:ibpreduction}

There are many publicly available codes that can automatize the procedure of the IBP reduction, but none of those tools can deal with integrals containing nonlinear propagators such as eq.\,\eqref{eq:nonlinprop} out-of-the box. Nevertheless, the general method of the IBP reduction is perfectly applicable to those integrals.
In this sense, we are facing a technical, rather than a conceptual limitation. To overcome this difficulty we split the reduction into two steps, which can be performed using different tools. In the first step we derive the IBP equations for each topology in the usual way by differentiating with respect to the loop momenta. This also applies to the nonlinear propagator. However, the so obtained system of equations turns out to be incomplete, since we miss a relation that would allow us to lower the integer power of the nonlinear propagator. Therefore, we need to augment the system by adding the relation
\begin{equation}
 \left ( 2 z \, p_a \cdot Q \, p_b \cdot Q - p_a \cdot p_b \right ) \left[ \delta(\mathcal{K}_{ab}(z)) \right ]^{j}   =\left[ \delta(\mathcal{K}_{ab}(z)) \right ]^{j-1}, \label{Eq:ibp-extra}
\end{equation}
with
\begin{equation}
\delta(\mathcal{K}_{ab}(z)) \equiv \frac{1}{2 \pi i} \left ( \frac{1}{2 z \, p_a \cdot Q \, p_b \cdot Q - p_a \cdot p_b - i \varepsilon}  - \frac{1}{2 z \, p_a \cdot Q \, p_b \cdot Q - p_a \cdot p_b + i \varepsilon} \right ).
\end{equation}
Once a solvable system of IBP equations has been obtained for each topology, we can proceed to the second step, where the reduction is performed by explicitly solving these equations for the relevant loop integrals using the Laporta algorithm~\cite{Laporta:2001dd}. Notice that this step can be performed without any knowledge about the explicit form of the propagators in the original topologies. Apart from the IBP equations and the list of the loop integrals we merely need to specify the positions of the cut propagators.

For the technical realization of the procedure described above, the publicly available tools \textsc{LiteRed}~\cite{Lee:2012cn} and \textsc{FIRE} 5~\cite{Smirnov:2014hma} were used to perform the steps one and two respectively. Although \textsc{LiteRed} actually does not support bases with nonlinear propagators, with a simple trick it can be nonetheless used to derive the corresponding IBP equations. The basic idea is to first create a new basis that omits  the nonlinear propagator, so that the \texttt{NewBasis} command can be evaluated without errors. The nonlinear propagator can be then added by directly modifying the list of the propagators \texttt{Ds} and the scalar product replacement rules \texttt{Toj}. These two modifications are sufficient to successfully run the \texttt{GenerateIBP} command of \textsc{LiteRed} and therefore to obtain the IBP equations for the given topology.
The extra relation from eq.\,\eqref{Eq:ibp-extra} can be easily added to the existing set of equations by applying the command \texttt{Toj} to a product of the inverse nonlinear propagator and a \texttt{j}-integral with symbolic indices.
Once we have the full set of valid IBP equations, it is a trivial exercise to convert them into the \textsc{FIRE} notation and save the result to a text file that will be later used by our \textsc{FIRE} code.

As far as \textsc{FIRE} is concerned, no additional tricks or modifications are required to use as it is a pure IBP solver. 
In order to run the C++ version of \textsc{FIRE} on the given set of loop integrals we need to generate the so-called \texttt{start}-files that contain all the relevant information about the current topology. The standard way to generate a \texttt{start}-file is to enter the basis by specifying the propagators, the loop and the external momenta. However, it is also possible to bypass this step and to enter only the corresponding IBP equations. This is done by assigning the set of the IBP equations  to the \textsc{FIRE} variable \texttt{startinglist}. We would like to stress that this working mode of \textsc{FIRE} is well documented (\cf section 4.1 of~\cite{Smirnov:2014hma}) and is supported since very early versions of the program. This feature obviously makes \textsc{FIRE} an ideal tool for our purposes.\footnote{The newest version of \textsc{KIRA}~\cite{Maierhoefer:2017hyi} also features the ability to solve custom systems of IBP equations. This functionality became publicly available only after this calculation has already been completed.} After having marked the cut propagators via the \texttt{RESTRICTIONS} variable, we can proceed with the commands \texttt{Prepare} and \texttt{SaveStart} that generate the corresponding \texttt{start}-file and save it to the disk respectively. In the following we can run the C++ version of \textsc{FIRE} with the obtained \texttt{start}-file in the usual way, just as one would do it for a standard set of propagators.

\subsection{Calculation of the master integrals}

With the aid of \textsc{FIRE} we can successfully reduce the loop integrals in each topology to a small set of master integrals. For this particular calculation no further steps are required, as, upon some loop momentum shifts  and renamings, all of the resulting master integrals can be related to the masters that were already computed in~\cite{Dixon:2018qgp}. However, since the explicit evaluation of these integrals requires some labor and due to the fact that~\cite{Dixon:2018qgp} contains almost no technical details regarding this step, we would like to provide more explicit explanations in this work.

In section \ref{sec:outline} we have already voiced our preference for solving the master integrals using the method of differential equations. We can regard the application of this technique as a two step process. First, we need to turn each system of differential equations into a canonical form~\cite{Henn:2013pwa}, provided that such a form exists.
The algorithms of Lee~\cite{Lee:2014ioa} or Meyer~\cite{Meyer:2016slj,Meyer:2018feh} can be used to construct the corresponding transformation matrix iteratively. Once the canonical form has been obtained, we can trivially determine the loop integrals at arbitrary order in $\varepsilon$ up to a set of unknown boundary constants.
To determine those constants we need to find suitable boundary conditions, which, depending on the process of interest, may turn out to be very challenging.

To implement the first step of the method, we can again use the tricks from section \ref{sec:ibpreduction} and employ \textsc{LiteRed} to differentiate with respect to $z$ via the command \texttt{Dinv}. The resulting loop integrals are then passed to \textsc{FIRE}. After another IBP reduction we obtain a closed system of differential equations in $z$. The search for the canonical form can be automatized using \textsc{Fuchsia}~\cite{Gituliar:2017vzm}, although other publicly available codes such as \textsc{Epsilon}~\cite{Prausa:2017ltv} or \textsc{CANONICA}~\cite{Meyer:2017joq} are also suitable for our purposes.

Some of our topologies can be converted into the canonical form only after performing a nonrational transformation of $z$, which is not automatically determined by the public \textsc{Python} version of \textsc{Fuchsia}\footnote{ It is worth noting that in the meantime the development C++ version of \textsc{Fuchsia} offers a helper tool that can suggest a suitable nonrational transformation, based on~\cite{Lee:2017oca}.}. Luckily, in our case the required transformations can be directly inferred from the analytic result for the standard EEC in $\mathcal{N}=4$ SYM~\cite{Belitsky:2013ofa}. The functional dependence of the classical polylogarihtms appearing in the final result on $\sqrt{z}$ readily suggests what kind of  transformations are required here. Consequently, we find that using one of the two transformations
\begin{align*}
z \to y^2, \quad z \to 1/(1-y^2)
\end{align*}
all systems of equations can be straightforwardly converted into the canonical form. The solutions of canonical basis can be easily written down iteratively order by order in $\epsilon$, all in terms of harmonic polylogarithms (HPLs)~\cite{Remiddi:1999ew} with argument $z $ or $y$, which can be conveniently manipulated using the \textsc{Mathematica} package \textsc{HPL}~\cite{Maitre:2005uu}. 

Then we can proceed to the second step, \ie the determination of the boundary constants. The difficulty to find suitable boundary conditions for the EEC observable makes this step the most time consuming part of the whole calculation. This is mainly due to the fact that the fixed order result diverges for $z\to0$ and $z\to1$ so that we cannot impose any regularity conditions in these limits.

Therefore, we have to use several different methods to determine the values of the integration constants up to $\mathcal{O}(\varepsilon^0)$ which is required for the NLO result. First of all, in the collinear limit $z\to0$ we can predict the power of the leading $1/z$ singularity by applying collinear power counting to each of the loop integrals. Working with light-cone coordinate $p = (p^0 + p^3, p^0 - p^3, p_\perp)$, when two of the four partons are collinear ($p_a \sim p_b \sim (\lambda^2, 1, \lambda) Q$ with $\lambda \ll 1$), the other two can be both anticollinear $p_c \sim p_d \sim (1, \lambda^2, \lambda) Q$, both hard $p_c \sim p_d \sim (1, 1, 1) Q$, anticollinear and collinear, or anticollinear and hard. Therefore, for each loop integral and each region we can assign a definite scaling in $\lambda$ to $z$, the measure and the occurring scalar products. For practical purposes, it is most convenient to work with the symmetric  parametrization of the 4-particle phase space from~\cite{Gehrmann-DeRidder:2003pne}. When the integral expanded around $z=0$ contains stronger poles than predicted by the power counting, the integration constants of those terms must be fixed in such a way, that these contributions vanish. Another requirement we impose is that the master integrals which are pure functions of uniform transcendental weight before converting to canonical basis must vanish in the unphysical limit $z \to \infty$. While the physical values of $z$ lie within the region $0\leq z \leq 1$, we can always perform an analytic continuation to the whole $z$ plane.

The most powerful but also most time-consuming procedure to determine the integration constants involves matching to the inclusive 4-particle phase-space master integrals from~\cite{Gehrmann-DeRidder:2003pne}. Our starting point is the obvious  identity
\begin{equation}
\hat{z}^n_{ab} (1 - \hat{z}_{ab})^m = \int_0^1 d z \, z^n (1-z)^m \, 2 p_a \cdot Q \, p_b \cdot Q \, \delta (2 z \, p_a \cdot Q \, p_b \cdot Q - p_a \cdot p_b), \label{Eq:z-identity}
\end{equation}
with $\hat{z}_{ab} = p_a \cdot p_b \, / (2 \, p_a \cdot Q \, p_b \cdot Q)$, where $m$ and $n$ are some nonnegative integers and the Dirac delta corresponds to the nonlinear propagator from eq.\,\eqref{eq:nonlinprop} that appears in our master integrals. These integrals can be represented as
\begin{equation}
\int d \Phi^{(4)} \mathcal{I}(\{ p \}) \, \delta (2 z \, p_a \cdot Q \, p_b \cdot Q - p_a \cdot p_b),
\label{eq:sampleint}
\end{equation}
where $d \Phi^{(4)}$ denotes the 4-particle Lorentz-invariant phase space and $\mathcal{I}(\{ p \})$ is the part of the integrand that does not depend on $z$. Multiplying the integrand with $z^n (1-z)^m \, 2 p_a \cdot Q \, p_b \cdot Q$ from eq.\,\eqref{Eq:z-identity} and with $(p_a \cdot Q \, p_b \cdot Q)^{m+n}$ and integrating over $z$ from $0$ to $1$ we obtain the relation
\begin{align}
& \int d \Phi^{(4)} \hat{z}^n_{ab} (1 - \hat{z}_{ab})^m (p_a \cdot Q \, p_b \cdot Q)^{m+n} \, \mathcal{I}(\{ p \}) \label{Eq:matching-bc} \\
& = \int_0^1 d z \, z^n (1-z)^m \int d \Phi^{(4)} \mathcal{I}(\{ p \}) 2 (p_a \cdot Q \, p_b \cdot Q)^{m+n+1} \, \delta (2 z \, p_a \cdot Q \, p_b \cdot Q - p_a \cdot p_b) \nonumber ,   
\end{align}
where $m$ and $n$ must be chosen such, that the integral over $z$ is convergent. The right-hand side of eq.\,\eqref{Eq:matching-bc} is evaluated as follows. First of all, we multiply the original integrand from eq.\,\eqref{eq:sampleint}  by $2 (p_a \cdot Q \, p_b \cdot Q)^{m+n+1}$ and reduce it to master integrals. All the resulting master integrals are already present in the corresponding system of differential equations, so that we know their solutions up to the unknown boundary constants. Then, we multiply the resulting linear combination of the master integrals by $z^m (1-z)^n$ and employ \textsc{HyperInt}~\cite{Panzer:2014caa} and \textsc{HPL}~\cite{Maitre:2005uu} to perform the integration over $z$. Now let us turn to the left-hand side of eq.\,\eqref{Eq:matching-bc}, where the loop integrals do not depend on $z$ and contain no nonlinear propagators. Hence, it is trivial to reduce those integrals to the master integrals of the inclusive 4-particle phase space, which are known analytically since long time~\cite{Gehrmann-DeRidder:2003pne}. Equating both sides of eq.\,\eqref{Eq:matching-bc} provides us with additional relations between the integration constants. Since each evaluation of the right-hand side of eq.\,\eqref{Eq:matching-bc} requires an IBP reduction of integrals containing the nonlinear propagator, the practical application of this method to fix the boundary constants is very time consuming. Moreover, to avoid spending too much on time on each single reduction, the integer values of $m$ and $n$ should be chosen sufficiently small, but still large enough to cure the divergences in $z \to 0$ and $z \to 1$ on the right-hand side of eq.\,\eqref{Eq:matching-bc}. Luckily, the values with $m,n \leq 1$ are enough for this calculation. We would also like to remark that this method bears similarity with the procedure employed in~\cite{Gituliar:2017umx}.
However, in our case it was much easier to apply, as our master integrals depend only on $z$, while the auxiliary master integrals from~\cite{Gituliar:2017umx} are functions of $z$ and the linearization parameter $x$. 

Finally, we also demand that after substituting all master integrals into the full NLO real correction, in the limit $z \to 0$ it may not diverge stronger than $1/z$~\cite{Konishi:1978yx,Richards:1982te}. When combined together, the above constraints are sufficient to determine all the required boundary constants.

\subsection{Real-virtual correction to Higgs EEC}

In the previous sections we were mainly concerned with the double-real corrections to the Higgs EEC. While this is undoubtedly the most important and also the most complicated ingredient of our analytic calculation, to obtain the full NLO result we must also include the real-virtual corrections. 

The evaluation of these corrections does not pose any difficulties. The  matrix element squared for the corresponding loop diagrams is currently known analytically at two loops, for both the dimension $5$ effective operator~\cite{Gehrmann:2011aa} and dimension $7$ effective operators~\cite{Jin:2018fak}. As a cross-check, we have explicitly recomputed the 1-loop contributions to the subprocesses $H \to ggg$ and $H \to q \bar{q} g$. The amplitudes were generated with \textsc{FeynArts} and evaluated using \textsc{FeynCalc}. To calculate the 1-loop integrals analytically we employed the \textsc{FeynHelpers}~\cite{Shtabovenko:2016whf}  interface between \textsc{FeynCalc} and \textsc{Package-X}~\cite{Patel:2015tea,Patel:2016fam}.

To carry out the renormalization, we included the counter terms for the dimension 5 effective operator to our \textsc{FeynRules} HEFT model and generated the corresponding counter term diagrams using \textsc{FeynArts}. Using the ability of \textsc{FeynCalc} and \textsc{Package-X} to explicitly distinguish between UV and IR poles at 1-loop, we have verified that
the inclusion of the counter term diagrams removes all UV poles in the real-virtual corrections, leaving us with IR poles only.

 The obtained results agree with the expressions arising from the analytic continuation and proper UV renormalization of the real-virtual contributions from~\cite{Kilgore:2013gba}. The final integrals over the massless 3-particle phase space can be carried out directly using the \textsc{HyperInt}~\cite{Panzer:2014caa} package. The still present IR poles in real-virtual cancel against the remaining IR poles in the double-real piece. As expected, the final result for the Higgs EEC is manifestly finite.

\section{Full analytic results}
\label{sec:fullres}
\begin{figure}[ht]
\centering
\includegraphics[width=0.9\textwidth,clip]{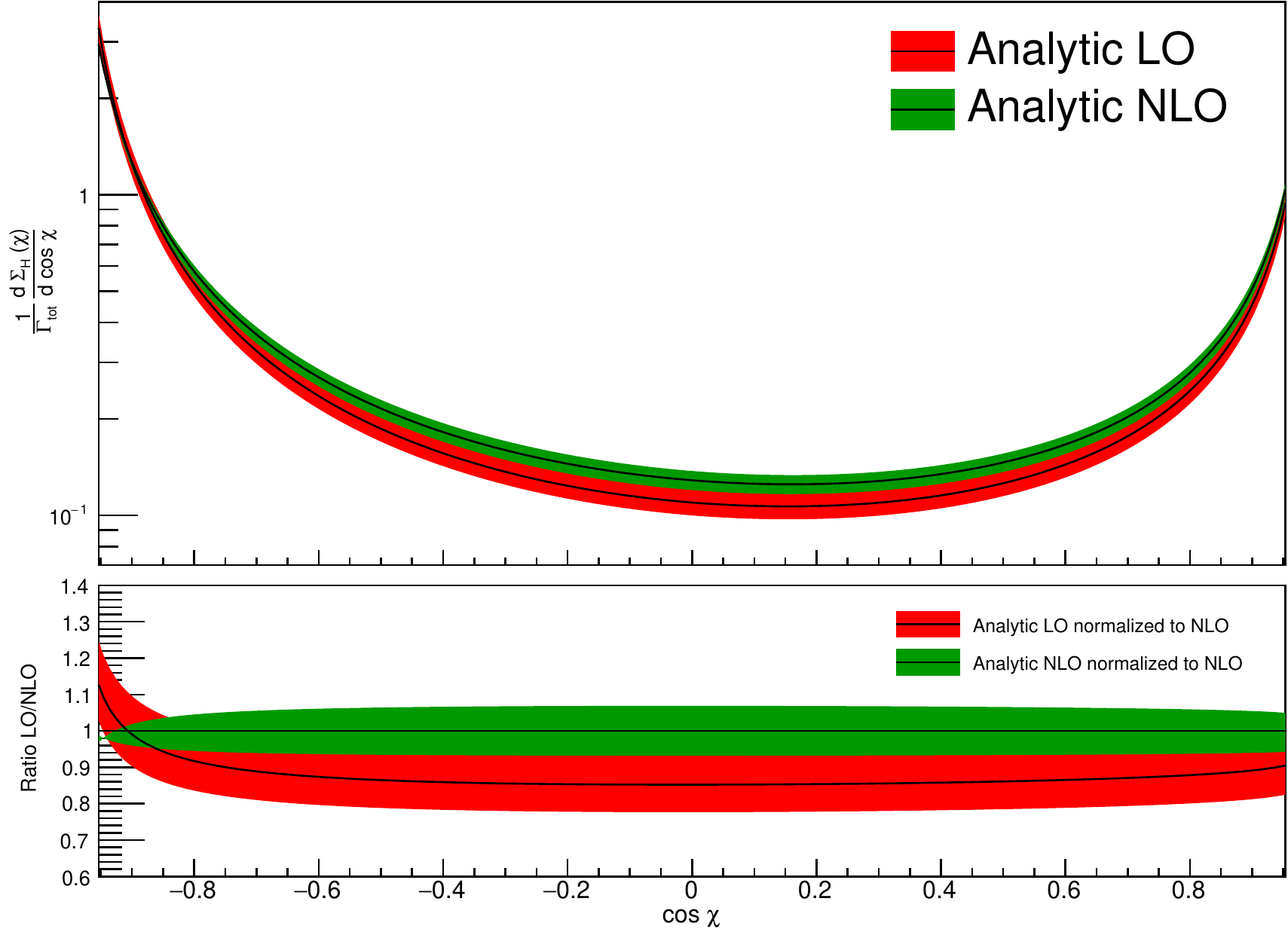}
\caption{Analytic fixed-order results for Higgs EEC at LO (lower curve) and NLO (upper curve) in the Higgs EFT. In both cases the solid black curves correspond to the central values, while the colored bands give the uncertainties from varying the renormalization scale $\mu$ between $m_H/2$ and $2 m_H$. We use $\mu = m_H$ as the central value. The number of flavors $N_f$ is set to 5 and the number of colors $N_c$ to 3.}
\label{fig:loNloPlot}
\end{figure}

The final analytic NLO result for the Higgs EEC can be written as 
\begin{align}
&   \frac{1}{\Gamma_{\textrm{tot}}} \frac{d\Sigma_H(\chi)}{d\cos\chi} =  \frac{1}{K(\mu)}  \left [
 \frac{\alpha_s(\mu)}{2\pi} A_H(z) + \left(\frac{\alpha_s(\mu)}{2 \pi}\right)^2  \left(3 \beta_0  \log \frac{\mu}{m_H} A_H(z) + B_H(z) \right)  + \mathcal{O} (\alpha_s^3) \right ],
 \label{eq:eecnlo}
\end{align}
with $A_H(z)$ and $B_H(z)$ being the LO and NLO coefficients respectively, while $\beta_0 = 11/3 C_A - 4/3 N_f T_f$. In QCD we have $C_A = N_c = 3$, $C_F = (N_c^2-1)/(2 N_c) = 4/3$ and $T_f = 1/2$, while $N_c$ denotes the number of colors. The overall prefactor $1/K$ arises from the normalization with respect to the total decay width for $H \to g g$ in HEFT, \cf eq.\,\eqref{eq:hdecaytot}. 
Figure \ref{fig:loNloPlot} shows the Higgs EEC at LO and NLO for $N_f=5$ with the corresponding uncertainties from varying the renormalization scale $\mu$ between $2 m_H$ and $m_H/2$. We set $m_H = 125.0\textrm{ GeV}$ and use the following values of the strong coupling $\alpha_s$ obtained with \textsc{RunDec}~\cite{Chetyrkin:2000yt,Herren:2017osy} at 4-loop accuracy
\begin{equation}
\alpha_s(m_H/2) = 0.125, \quad \alpha_s(m_H) = 0.113, \quad \quad \alpha_s(2 m_H) = 0.103.
\end{equation}

Before presenting the explicit result for the LO and NLO contributions let us, for the sake of clarity, decompose the corresponding coefficients $A_H(z)$ and  $B_H(z)$ into different color pieces that appear in the full result. For $A_H(z)$ we will use the subscript ``lc'' to denote the leading color contribution ($\sim N_c$). In the case of $B_H(z)$ ``lc'' stands for the component proportional to $N_c^2$, ``nlc'' for the next-to-leading color part ($\sim N_c$) and ``nnlc'' for the next-next-to-leading color piece ($\sim 1/N_c$). The components proportional to the number of flavors $N_f$ will be named accordingly. For LO we find
\begin{align}
A_H(z) & = C_A A_{H,\text{lc}}(z) + N_f T_f A_{H,N_f}(z)
\label{eq:ahzexpl}
\end{align}
with
\begin{subequations}
\begin{align}
A_{H,\text{lc}}(z) & = \frac{25 z^3-156 z^2+336 z-216}{12 (1-z) z^5}-\frac{\left(2 z^4-14 z^3+51 z^2-74 z+36\right)}{2 (1-z) z^6}\log (1-z) \,,\\ 
A_{H,N_f}(z) & = \frac{-25 z^3+201 z^2-390 z+216}{6  (1-z) z^5} + \frac{\left(-z^3+16 z^2-47 z+36\right) }{z^6} \log (1-z) \,.
\end{align}
\end{subequations}
The color decomposition of the NLO coefficient $B_H(z)$ reads
\begin{equation}
B_H(z) = C_A^2 B_{H,\text{lc}}(z) + C_A T_f N_f B_{H,\text{nlc}}(z)+ (C_A - 2 C_F) T_f N_f B_{H,\text{nnlc}}(z) + N_f^2 T_f^2 B_{H,N_f^2}(z),
\label{eq:bhzdecomp}
\end{equation}

It is interesting to observe that the Higgs EEC exhibits a much richer color structure as compared to the standard EEC. While the LO coefficient of the standard EEC is directly proportional to $C_F$, eq.\,\eqref{eq:ahzexpl} depends not only on $C_A$ but also on $N_f$. Notice that the $N_f$-piece in the standard EEC is an NLO effect, while in the Higgs EEC it appears already at LO. Furthermore, unlike Eq\,\eqref{eq:bhzdecomp} the $B(z)$ coefficient of the standard EEC contains no terms proportional to $N_f^2$. These differences can be largely attributed to the fact that the Higgs EEC is a gluon-initiated observable, while the standard EEC is a quark-initiated quantity. For example, the splitting of a gluon into a quark-antiquark pair is a LO effect in the Higgs EEC, but for the standard EEC it can occur only at NLO and beyond.

The mathematical structure of the explicit expressions for the color coefficients from eq.\,\eqref{eq:bhzdecomp} is very similar to what has already been observed in the analytic NLO result for the standard EEC. Each term is essentially a product of a rational function of polynomials in $z$ multiplying a building block function $g_i^{(j)}$ of pure transcendental weight $j$. These functions are
\begin{align}
  g_1^{(1)} &= \log (1-z)\,, \nonumber \\
  g_2^{(1)} &=  \log (z)\,,  \nonumber \\
  g_1^{(2)} &= 2 (\text{Li}_2(z)+\zeta_2)+\log ^2(1-z)\,, 
\nonumber\\
g_2^{(2)} & = \text{Li}_2(1-z)-\text{Li}_2(z)\,, \nonumber \\
g_3^{(2)} &= - 2 \, \text{Li}_2\left(-\sqrt{z}\right)
+ 2 \, \text{Li}_2\left(\sqrt{z}\right)
+ \log\left(\frac{1-\sqrt{z}}{1+\sqrt{z}}\right) \log (z) \,,
\nonumber\\
g_4^{(2)} &= \zeta_2 \,, \nonumber \\
g_1^{(3)}  & = -6
\left[ \text{Li}_3\left(-\frac{z}{1-z}\right)-\zeta_3 \right]
- \log \left(\frac{z}{1-z}\right)
 \left(2 (\text{Li}_2(z)+\zeta_2)+\log^2(1-z)\right)
\,,
\nonumber\\
g_2^{(3)} & =  -12
\left[ \text{Li}_3(z)+\text{Li}_3\left(-\frac{z}{1-z}\right) \right]
+ 6 \, \text{Li}_2(z) \log(1-z) + \log^3(1-z) \,,
\nonumber\\
g_3^{(3)} & = 6 \log(1-z) \, (\text{Li}_2(z)-\zeta_2)
- 12 \, \text{Li}_3(z) + \log^3(1-z) \,, \nonumber\\
g_4^{(3)} &= \text{Li}_3\left(-\frac{z}{1-z}\right)
                - 3 \, \zeta_2 \log(z) + 8 \, \zeta_3 \,,\nonumber\\
g_5^{(3)} &=  
- 8 \left[ \text{Li}_3\left(-\frac{\sqrt{z}}{1-\sqrt{z}}\right)
+ \text{Li}_3\left(\frac{\sqrt{z}}{1+\sqrt{z}}\right) \right]
+ 2 \text{Li}_3\left(-\frac{z}{1-z}\right)
+ 4 \zeta_2 \log (1-z) \nonumber \\ 
& +\log \left(\frac{1-z}{z}\right)
     \log^2\left(\frac{1+\sqrt{z}}{1-\sqrt{z}}\right) \,.
\label{eq:gdef}
\end{align}
All these functions have only $z$ and $\sqrt{z}$ dependence, but intermediate results do have explict dependence on $i \sqrt{z}/\sqrt{1-z}$, which cancels out when combining real-virtual contributions with $q\bar{q}gg $ cut or $gggg$ cut.
 This is a strong check of the correctness of our final results.

The numerator of the rational function is an univariate polynomial with the highest power of $z$ being 8, while the denominator is a function of type $(1-z)^m z^k$ with an integer $m$ between 0 and 1, $k$ between 0 and 6. The only exception to this rule arises from the single terms present in $B_{H,\text{lc}}(z)$, $B_{H,\text{nlc}}(z)$ and $B_{H,\text{nnlc}}(z)$ which
explicitly depend on $\sqrt{z}$ through $g_3^{(2)}$ and the corresponding coefficients. The products of those with $g_3^{(2)}$ are symmetric with respect to $\sqrt{z} \to -\sqrt{z}$, a property that has already been observed in the standard EEC result~\cite{Dixon:2018qgp} and the result in $\mathcal{N}=4$ SYM~\cite{Belitsky:2013ofa}. 

It is worth noting that $B_{H,N_f^2}(z)$, which is the simplest piece of $B_H(z)$, turns out to be much simpler than the $B_{N_f}(z)$ part of the QCD result. While the latter contains one weight 3 function and 
an explicit dependence on $\sqrt{z}$, the former depends only two weight 1 and three weight 2 functions and is free of square roots.

In the following we list the explicit values of the color components of $B_H(z)$.

\begin{subequations}
\begin{align}
& B_{H,\text{lc}}(z) = -\frac{{3240 z^6-3240 z^5+981 z^4-207539 z^3+1131821 z^2-2416929 z+1546086}}{8640  (1-z) z^5} \nonumber \\
&-\frac{{2160 z^8-3780 z^7+5640 z^6-3909 z^5+2317 z^4+12434 z^3-2958 z^2-36449 z+22565}}{1440 (1-z) z^6 } { g_1^{(1)}} \nonumber\\
&+\frac{{2160 z^7-2700 z^6+4560 z^5-975 z^4-13190 z^3+70367 z^2-151398 z+92556}}{1440 (1-z) z^5} { g_2^{(1)}} \nonumber \\ 
&+\frac{{ -168 z^6+353 z^5-605 z^4+3080 z^3-3860 z^2-1967 z+4047}}{240  (1-z) z^6} { g_1^{(2)}} \nonumber \\
&-\frac{{-180 z^7+90 z^6-330 z^5+75 z^4-460 z^3+3000 z^2-8860 z+7833 }}{120 z^6} { g_2^{(2)}} \nonumber\\
& -\frac{{1920 z^4+725 z^3+1789 z^2-640 z+960}}{960 z^{11/2}} { g_3^{(2)}} \nonumber \\
&  +\frac{{ -240 z^5+515 z^4-3130 z^3+7770 z^2-7303 z+1893 }}{60  (1-z) z^6 } { g_4^{(2)} } \nonumber\\
& -\frac{{ 3 z^4-6 z^3+9 z^2-10 z+3}}{4 (1-z) z} { g_1^{(3)}} -\frac{ { 2 z^6-z^5+7 z^4-44 z^3+156 z^2-224 z+109}}{12  (1-z) z^6} { g_2^{(3)}} \nonumber\\
& +\frac{1}{6 (1-z)} { g_3^{(3)}} +\frac{{ 1-2 z}}{2 (1-z) z} { g_4^{(3)}} +\frac{{ 2 z^5+z^4+2 z^2-z+1}}{4  z^6} { g_5^{(3)} },  \\
& B_{H,\text{nlc}}(z) = +\frac{1080 z^6-1080 z^5+219 z^4-111086 z^3+929054 z^2-1863951 z+1053294}{2160 (1-z) z^5} \nonumber  \\
& +\frac{720 z^8-1260 z^7+800 z^6-241 z^5-5602 z^4+18841 z^3+9973 z^2-54396 z+30145}{360 (1-z) z^6} g_1^{(1)}  \nonumber \\
& -\frac{ 1440 z^7-1800 z^6+880 z^5-110 z^4-19445 z^3+127791 z^2-230784 z+122328}{720 (1-z) z^5} g_2^{(1)} \nonumber \\
& -\frac{3 z^6-3 z^5-585 z^4+3220 z^3-4005 z^2-473 z+1923}{60 (1-z) z^6} g_1^{(2)}  \nonumber \\
& +\frac{ -120 z^7+60 z^6-40 z^5-20 z^4-480 z^3+5280 z^2-14135 z+10194}{60 z^6} g_2^{(2)} - \frac{3-25 z}{480 z^{7/2}} g_3^{(2)}  \nonumber \\
& -\frac{1630 z^4-12200 z^3+27425 z^2-23383 z+6348}{60 (1-z) z^6} g_4^{(2)} \nonumber \\
& -\frac{2 z^2-2 z+1}{2}  g_1^{(3)} +\frac{ (2-z) \left(z^2-9 z+10\right)}{z^6} g_2^{(3)},  \\ \nonumber \\
& B_{H,\text{nnlc}}(z) = +\frac{-360 z^6+360 z^5-z^4-3231 z^3+69389 z^2-158391 z+91874}{1440 (1-z) z^5} \nonumber \\
& +\frac{ 720 z^7-540 z^6+260 z^5-17 z^4+9434 z^3-28724 z^2+10742 z+14245 }{720 z^6} g_1^{(1)} \nonumber \\
& -\frac{ 720 z^6-180 z^5+260 z^4+205 z^3+9710 z^2-33534 z+31692 }{720 z^5} g_2^{(1)} \nonumber \\
& +\frac{ -6 z^5+10 z^4-760 z^3+3255 z^2-4810 z+2521 }{120 z^6} g_1^{(2)} + \frac{2 z^2-2 z+1 }{4}  g_1^{(3)} \nonumber \\
& +\frac{ 60 z^7-30 z^6+20 z^5+10 z^4-270 z^3+1935 z^2-4015 z+2521}{60 z^6} g_2^{(2)} \nonumber \\
& -\frac{ 25 z^3+88 z^2-160 z+240}{120 z^{11/2}} g_3^{(2)}+\frac{-6 z^3+35 z^2-52 z+22}{12 z^6} g_2^{(3)}  \nonumber \\
& -\frac{-1070 z^3+5250 z^2-8945 z+5042}{60 z^6} g_4^{(2)}  +\frac{z^2-2 z+2}{4 z^6} g_5^{(3)}, \\ \nonumber \\
& B_{H,N_f^2}(z) = -\frac{-623 z^3+6149 z^2-12633 z+7122}{54 (1-z) z^5} -\frac{2 \left(17 z^2-42 z+24\right)}{9 z^5} g_2^{(1)} \nonumber \\
& -\frac{ \left(107 z^4-974 z^3+2799 z^2-3118 z+1174\right)}{18 (1-z) z^6} g_1^{(1)}  +\frac{ \left(-5 z^3+60 z^2-159 z+116\right)}{3 z^6} g_1^{(2)} \nonumber \\
& +\frac{4  (1-z)^2 (4-z)}{3 z^6} g_2^{(2)}  -\frac{2  \left(-7 z^3+72 z^2-177 z+124\right)}{3 z^6} g_4^{(2)}.
\end{align}
\end{subequations}

\begin{figure}[ht]
\centering
\includegraphics[width=0.9\textwidth,clip]{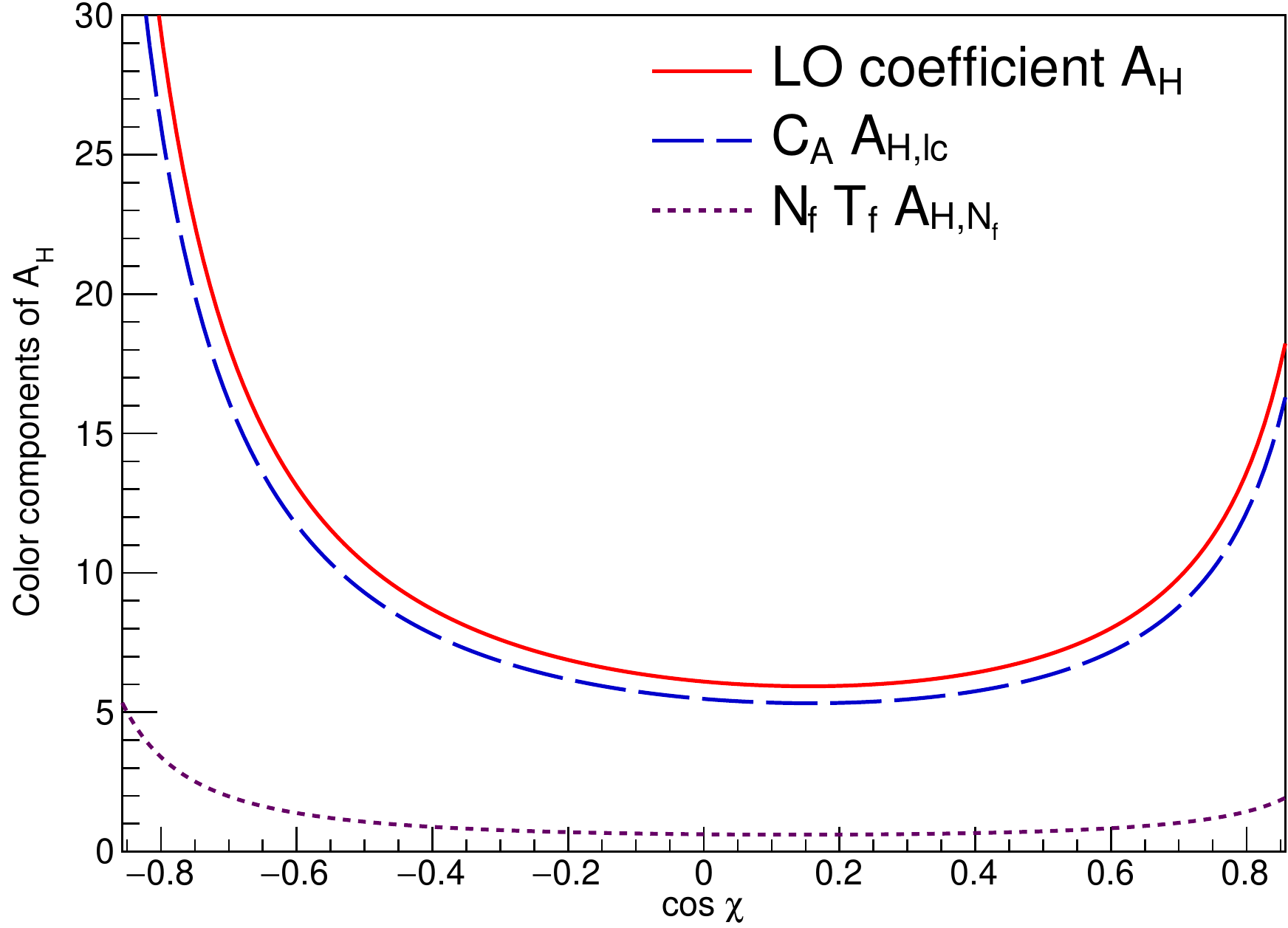}
\caption{LO coefficient $A_H$ and its color components $A_{H,\textrm{lc}}$ and
$A_{H,N_f}$ for $N_f =5$ and $N_c = 3$. Both components give positive contributions.}
\label{fig:ahplot}
\end{figure}

\begin{figure}[ht]
\centering
\includegraphics[width=0.9\textwidth,clip]{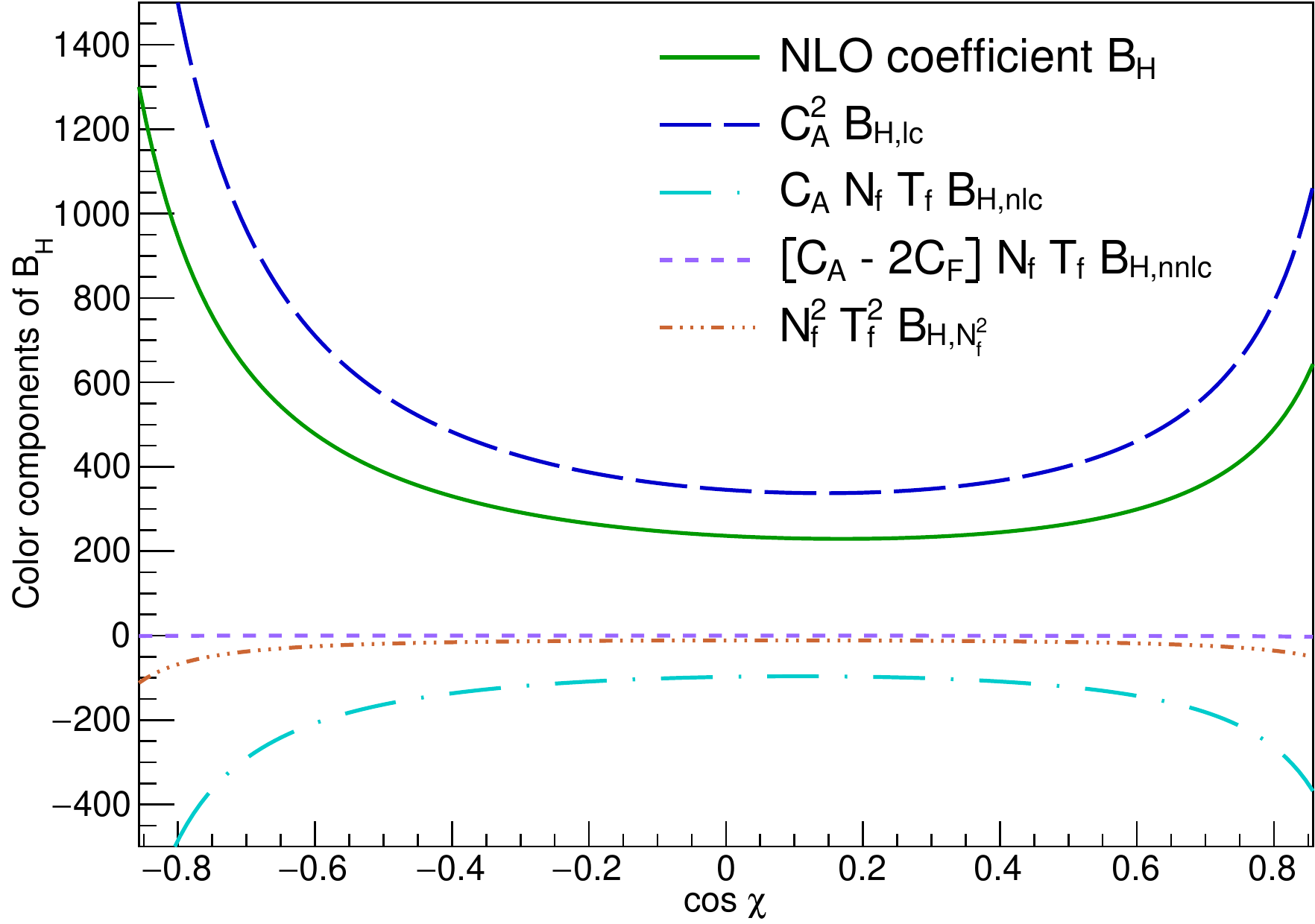}
\caption{NLO coefficient $B_H$ and its color components $B_{H,\textrm{lc}}$, $B_{H,\textrm{nlc}}$, $B_{H,\textrm{nnlc}}$ and $B_{H,N_f^2}$ for $N_f =5$ and $N_c = 3$. Notice that only the contribution of $B_{H,\textrm{lc}}$ is positive, while the three other components contribute negatively.}
\label{fig:bhplot}
\end{figure}

To assess the relative importance of the different color components for the final result, we plot $A_H(z)$ and $B_H(z)$ together with their color coefficients in figure \ref{fig:ahplot} and figure \ref{fig:bhplot} respectively.

\section{Asymptotics}
\label{sec:asymptotics}

Let us now explore the asymptotics of the Higgs EEC by expanding the full NLO result in the back-to-back and collinear limits. For the limit $z \to 0$ expanded up to $\mathcal{O}(z)$ we find

\begin{subequations}
\begin{align}
A_H(z)  & =  \frac{1}{z} \left[\frac{7 C_A}{20}+ \frac{N_f T_f}{20}\right]+\frac{59 C_A}{120}+\frac{N_f T_f}{15} + \mathcal{O}(z) \,, \\ \nonumber \\ 
B_H(z)  & = \frac{1}{z} \biggl[\log (z) \left(-\frac{91 C_A^2}{600} + \frac{89 C_A N_f T_f}{200} -\frac{7 C_F N_f T_f}{40} +\frac{N_f^2 T_f^2}{15} \right) \nonumber \\
& + C_A^2 \left(-\frac{\zeta_3}{2} + \frac{97 \zeta_2}{60}+\frac{138427}{27000}\right) 
 +C_A N_f T_f \left(-\frac{7 \zeta_2}{30}-\frac{201371}{108000}\right) +\frac{9 C_F N_f T_f}{400} \nonumber \\
 & -\frac{43 N_f^2 T_f^2}{120} \biggr ] +\log (z) \biggl[C_A^2 \left(-2 \zeta_2 +\frac{9221}{3150}\right)+C_A N_f T_f \left(\frac{\zeta _2}{2}-\frac{8999}{25200}\right) \nonumber \\ 
 & +C_F N_f T_f \left(\zeta_2-\frac{3163}{1800}\right) +\frac{2}{45} N_f^2 T_f^2\biggr] + C_A^2 \left(\frac{13 \zeta_3}{2}-\frac{47 \zeta _2}{15}+\frac{79860499}{10584000}\right) \nonumber \\
& +C_A N_f T_f \left(-\frac{3 \zeta_3}{2}+\frac{13 \zeta _2}{20}-\frac{2703293}{1176000}\right) +C_F N_f T_f \left(-3 \zeta _3+\frac{41 \zeta _2}{30}+\frac{125143}{108000}\right) \nonumber \\
& -\frac{2207 N_f^2 T_f^2}{5400} + \mathcal{O}(z).
\end{align}
\end{subequations}

The other limit, $z \to 1$ corresponds to the back-to-back limit. In this limit, the large logarithms originate from soft and collinear radiations. The resummation of large logarithms in this limit is well understood, and is closely related to various transverse-momentum dependent distribution~\cite{Collins:1981uk,Dokshitzer:1999sh,Moult:2018jzp,Gao:2019ojf}. The expansion up to $\mathcal{O}(1-z)$ yields

\begin{subequations}
\begin{align}
A_H(z) & =\frac{1}{1-z} \biggl [-\frac{1}{2} C_A \log (1-z)-\frac{11 C_A}{12}+\frac{N_f T_f}{3}\biggr ]+ \left(-6 C_A + 4 N_f T_f\right) \log (1-z) \nonumber \\
& -\frac{77 C_A}{6}+\frac{73 N_f T_f}{6} + \mathcal{O}(1-z), \\ \nonumber \\ 
B_H(z) & = \frac{1}{1-z} \biggl[\frac{C_A^2}{2}  \log ^3(1-z)+ \left(\frac{11 C_A^2}{3}-\frac{4 C_A N_f T_f}{3} \right) \log ^2(1-z)
\nonumber \\
& + \log (1-z) \left(C_A^2 \left(\frac{3 \zeta _2}{2}+\frac{11}{8}\right)  -\frac{17}{6} C_A N_f T_f+ \frac{2}{3} N_f^2 T_f^2\right) +C_A^2 \left(\frac{\zeta_3}{2}+\frac{77 \zeta_2}{12}-\frac{907}{144}\right)
\nonumber \\
&+ C_A N_f T_f \left(-\frac{7 \zeta_2}{3}+\frac{233}{72}\right) +\frac{C_F N_f T_f}{2} -\frac{5 N_f^2 T_f^2}{18} \biggr ] 
\nonumber \\
&+\log (1-z) \biggl [C_A^2 \left(16 \zeta _2+\frac{415}{18}\right) + C_A N_f T_f \left(-\frac{23 \zeta _2}{2}-\frac{211}{18}\right) +C_F N_f T_f \left(\zeta_2-17\right) \\
& +\frac{29 N_f^2 T_f^2}{9} \biggr]  +\log ^3(1-z) \left(\frac{10 C_A^2}{3}-\frac{23 C_A N_f T_f}{12} -\frac{C_F N_f T_f}{6} \right)
\nonumber \\
& +\log ^2(1-z) \left(\frac{757 C_A^2}{24} -\frac{307 C_A N_f T_f}{12} -\frac{7 C_F N_f T_f}{2} +4 N_f^2 T_f^2\right)
\nonumber \\
& +C_A^2 \left(\frac{155 \zeta_3}{4} +15 \zeta _2 \log (2) + \frac{7001 \zeta _2}{96}-\frac{7115}{54}\right) \nonumber \\
& +C_A N_f T_f \left(-\frac{105 \zeta_3}{4} +3 \zeta_2 \log (2) -\frac{1387 \zeta_2}{16}+\frac{21163}{108}\right)
\nonumber \\
& +C_F N_f T_f \left(- \frac{3 \zeta_3}{2} -6 \zeta _2 \log (2) + \frac{151 \zeta_2}{12}-\frac{929}{24}\right)+N_f^2 T_f^2 \left(8 \zeta _2-\frac{2279}{54}\right) + \mathcal{O}(1-z).
\end{align}
\end{subequations}
Note that the appearance of $\zeta_2 \log(2)$ in the constant term at next-to-leading power, which comes from both $B_{H,\text{lc}}$ and $B_{H,\text{nnlc}}$, is different from standard EEC~\cite{Dixon:2018qgp}, where it originates solely from the next-to-leading color part. We note that the leading power terms in the $z \to 1$ limit are in full agreement with the factorization prediction based on the formalism in ref.\,\cite{Moult:2018jzp}.\footnote{Private communication with Anjie Gao.} The subleading power corrections of the Higgs EEC are closely related to the perturbative power corrections for Drell-Yan/Higgs $p_T$ distribution at hadron collider. In the latter case, perturbative power corrections in the presence of rapidity divergence~\cite{Ebert:2018gsn} has been studied very recently. We expect the same method can be used to compute the perturbative power corrections for EEC. 

\section{Estimates of nonperturbative corrections}
\label{sec:numerics}

As every QCD observable, Higgs EEC necessarily contains nonperturbative corrections to the result obtained in pQCD. Even though they are considered to be suppressed as the inverse of the relevant energy scale (in our case $m_H$), these contributions are necessary for a meaningful comparison between theoretical predictions and experimental data. In practice, the size of the nonperturbative effects can be estimated using a suitable model (\eg DWM~\cite{Dokshitzer:1999sh}) or by simulating parton shower and hadronization with software tools such as \textsc{Pythia}~\cite{Sjostrand:2014zea}. The latter approach is also what we choose to assess the influence of the nonperturbative corrections on the Higgs EEC. 

In our \textsc{Pythia} setup we consider the process $e^+ e^- \to H \to gg$ at $\sqrt{s} = m_H$ and generate 5000 events, which, according to~\cite{An:2018dwb} roughly corresponds to what CEPC expects to collect in the $H \to gg$ decay channel over a data taking period of 7 years. The tiny size of the Higgs to electrons coupling is irrelevant here, as the Higgs EEC is obtained from the decay of the Higgs particle at rest and does not depend on its production mechanism.

\textsc{Pythia} itself implements only the LO hard matrix element for $H \to gg$ and employs parton shower to approximate higher order corrections. This is sufficient to simulate gluon-initiated $H \to q \bar{q} g$ final states, but turns out to be problematic in the case of $H \to g g g$. The reason for this is that the gluon splitting is not the only contribution to the 3 gluon final state. The other contribution arises directly from the 3-gluon-Higgs interaction vertex and is not negligible. However, the latter is not implemented in the current version of \textsc{Pythia}, so that the simulation of the $H \to g g g$ final state is obviously incomplete.

In principle, one could overcome this issue by matching the current \textsc{Pythia} two jet + parton shower simulation to the full LO matrix element that incorporates both 2-gluon and 3-gluon vertices. However, since we are mainly interested in a \textit{qualitative} comparison between \textsc{Pythia} and our analytic result, we choose not to go this route and content ourselves with what is already implemented in the code of \textsc{Pythia}.

Despite of these shortcomings, we believe that the comparison of our NLO results to the \textsc{Pythia} predictions is still interesting, especially in view of the lack of any experimental data for the Higgs EEC event shape observable. A more careful and meaningful comparison to a proper numerical simulation that uses the NLO hard matrix element shall be addressed in a future work.

  The Higgs EEC distribution is calculated according to~\cite{Abreu:1996na}
\begin{equation}
\Sigma_H  (\chi) = \frac{1}{N_\textrm{events}} \frac{1}{\Delta \cos \chi } \sum_{N_\textrm{events}} \sum_{a,b} \frac{E_a E_b}{E_{\textrm{vis}}^2} \Theta \left ( \Delta \cos \chi - |\cos \chi - \cos \chi_{ab}| \right ), \label{eq:eec}
\end{equation}
where $\chi_{ab}$ is the angle between the directions of the 3-momenta of the final state particles $a$ and $b$, $\Delta \cos \chi$ is the histogram bin width and $\cos \chi$ is the lower edge of the bin. The overall normalization by the total number of events $N_\textrm{events}$ and the bin width ensures that the area under the histogram is unity.

\begin{figure}[ht]
\centering
\includegraphics[width=0.9\textwidth,clip]{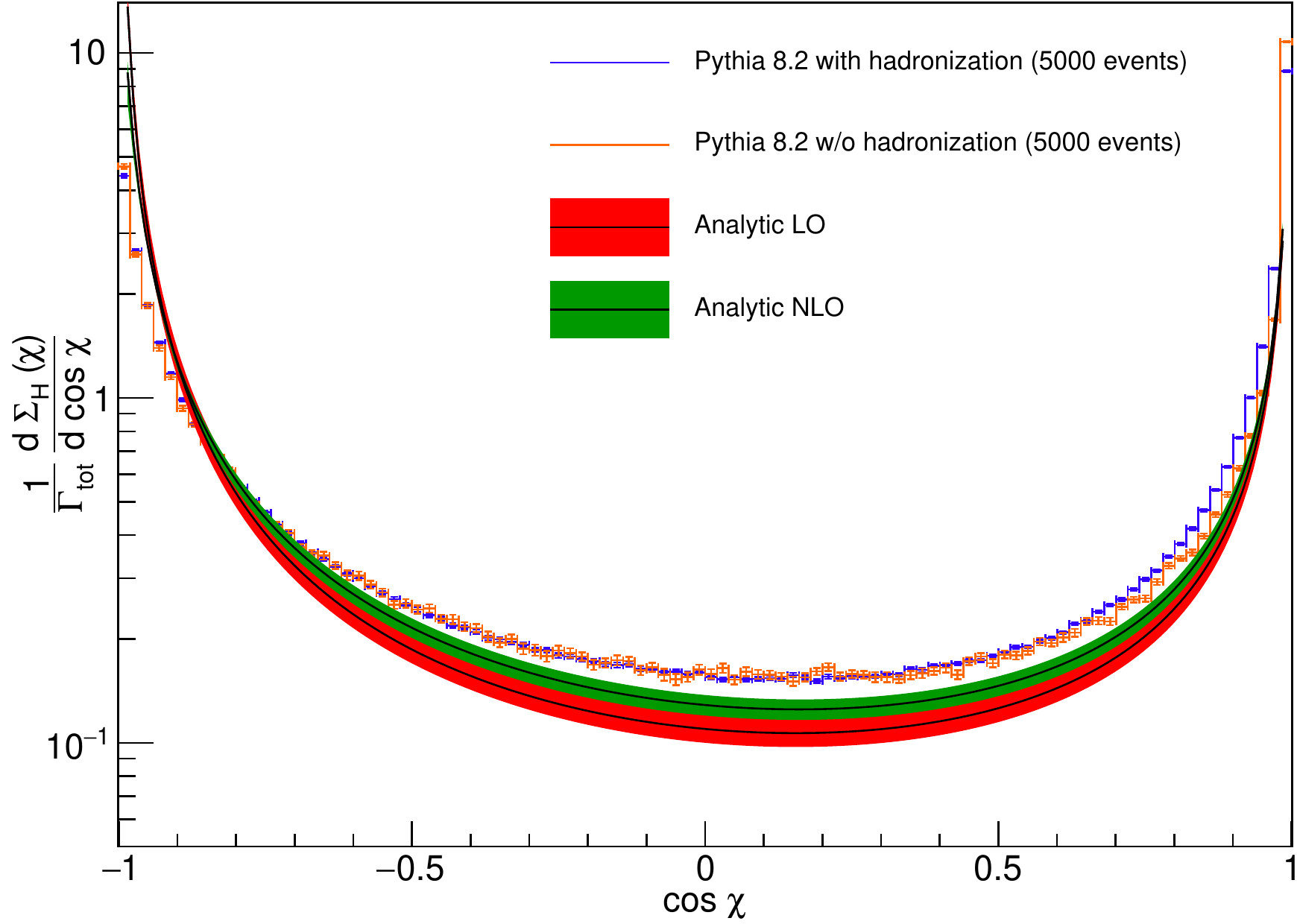}
\caption{Comparison of a \textsc{Pythia} simulation for Higgs EEC to the analytic LO and NLO results from eq.\,\eqref{eq:eecnlo}. Both \textsc{Pythia} curves contain contributions from self-correlations, which are not included in the analytic result. The area under both \textsc{Pythia} curves is unity. Omitting the self-correlations decreases the area under the \textsc{Pythia} curve with haronization to $0.96$, while the area under the curve without hadronization becomes $0.88$. 
Adding self-correlations only increases the number of entries in the very last bin in the collinear ($\cos \chi \approx 1$) region, while the rest of the curve remains unchanged.}
\label{fig:pythiaanalytic}
\end{figure}

We employ \textsc{Pythia} 8.2 and \textsc{ROOT} 6.14~\cite{Brun:1997pa} to simulate gluon-initiated Higgs 
decays both to hadrons and to partons. In the latter case the hadronization effects are disabled via the master switch \texttt{HadronLevel:Hadronize = 0}. Only statistical uncertainties are included. The corresponding plot is shown in figure \ref{fig:pythiaanalytic}. The discrepancy between the \textsc{Pythia} data and the NLO result can be attributed to the missing  $H \to ggg$ vertex in \textsc{Pythia} hard matrix element, sizeable NLO corrections and nonperturbative effects.

For simplicity, we choose to ignore the systematic errors of this simulation, so that the displayed error bars stem only from statistical uncertainties. Since a single event with more than 2 finite state particles generates multiple histogram entries, the errors in different angular bins are statistically correlated. Thus, it does not seem appropriate to calculate the errors assuming that the simulated data obeys an uncorrelated Poisson distribution. Instead, we adopt the approach that was used by the TOPAZ experiment when measuring EEC at the TRISTAN collider \cite{Adachi:1989ej}. To be more specific, for each \textsc{Pythia} configuration we generate 50 additional samples with the same settings and the same number of events but different random seeds. Then, for each bin $i$ we can calculate the standard deviation $\sigma_i$ via
\begin{equation}
\sigma_i = \sqrt{ \frac{1}{n} \sum_{j=1}^n (x_j(i) - \mu_i)^2},
\end{equation}
where $x_j(i)$ is the content of the $i^\textrm{th}$ bin in the $j^\textrm{th}$ sample, $n=50$ and the mean for the
$i^\textrm{th}$ bin, $\mu_i$ is given by
\begin{equation}
\mu_i = \frac{1}{n} \sum_{j=1}^n x_j(i).
\end{equation}
Comparing the size of the errors in our simulations we observe the unphysical effect that the curve with hadronization seems to have smaller errors than the one without. This suggests that the systematic errors (especially for hadronization) might be much larger than the statistical uncertainties.  Looking only at the central values, it is interesting to observe that in the $\cos \chi$ region between $-0.5$ and $0.5$, both curves seem to lie very close to each other. However, given the possibility of underestimated uncertainties, we must abstain from deriving any conclusions on the size of the nonperturbative corrections.

\begin{figure}[ht]
\centering
\includegraphics[width=0.9\textwidth,clip]{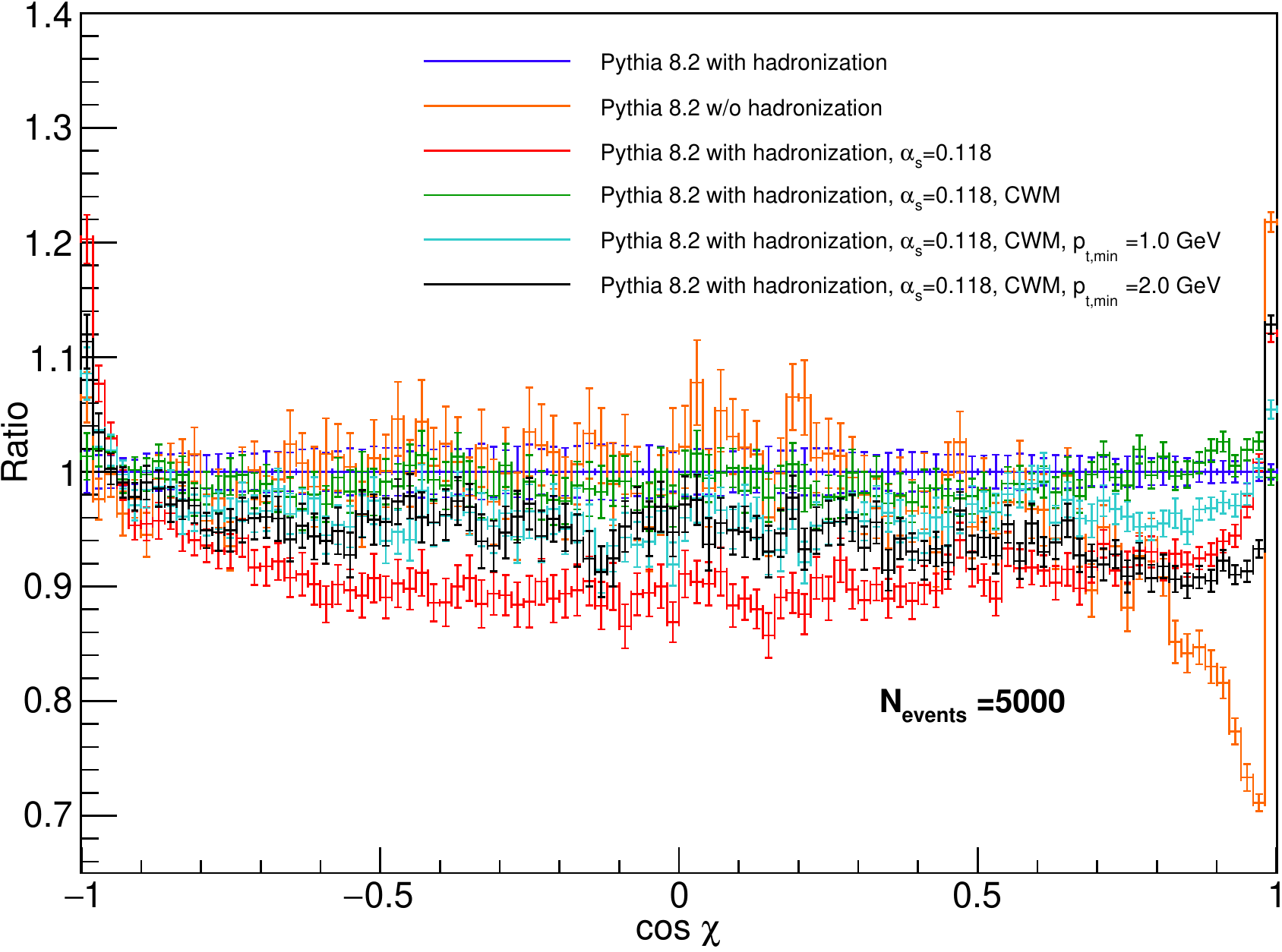}
\caption{Variation of some \textsc{Pythia} parameters that govern the hadronization. For the sake of clarity, all curves are normalized to the curve with hadronization generated using the standard settings (blue curve). In the central region the largest discrepancy to the standard settings arises when changing the effective value of $\alpha_s$ but keeping the CWM resummation disabled (red curve). Once the CWM resummation is activated (green, cyan and black curves) the obtained results become very similar to the default output.}
\label{fig:pythiaHadroCheck}
\end{figure}

To check whether our observation is a genuine \textsc{Pythia} effect or a possible artifact of an accidental fine-tuning, we varied several important parameters responsible for simulating the hadronization effects according to~\cite{Skands:2014pea}. In particular, we set the effective value of $\alpha_s (M_Z)$ in the final-state radiation to the physical value $0.118$ (\texttt{TimeShower:alphaSvalue = 0.118}), activated the CWM~\cite{Catani:1990rr} resummation of subleading terms using the 2-loop running of $\alpha_s$ (\texttt{TimeShower:alphaSuseCMW = on}, \texttt{TimeShower:alphaSorder = 2}) and varied the value of the IR shower cutoff (\texttt{TimeShower:pTmin}) by setting it to $1.0\textrm{ GeV}$ and $2.0\textrm{ GeV}$. As can be inferred from figure \ref{fig:pythiaHadroCheck}, the
effects of these changes in the parameter sets are rather mild and they seem to support 
the initial observation that, judging from the central values only, \textsc{Pythia} possibly indicates comparably small influence of the hadronization for $|\cos \chi| \leq 0.5$. Those effects, however, become more prominent in the collinear and back-to-back regions, $z \to 0$ and $z \to 1$ respectively. Of course, these naive comparisons should not be interpreted as a definitive statement on the size of the hadronization effects in the Higgs EEC.
To make such a statement one would need to perform a much more careful numerical simulation and error analysis. Here we merely describe our observations and speculate about
their possible interpretation.

Finally, it is tempting to attempt a fit of the analytic NLO result to the \textsc{Pythia} prediction in order to perform a toy determination of $\alpha_s$. Our motivation is clear: It is well known that the standard EEC measurements at LEP and other experiments were often used to determine the value of the strong coupling. Therefore, should a future lepton collider measure the $H \to gg$ channel with sufficient precision, it will not take long until experimentalists and theorists will attempt to extract the value of $\alpha_s$ from the Higgs EEC measurements. The purpose of our toy fit is to attract the attention of the high energy physics community to this scenario and to ensure that when the real data becomes available, all other ingredients for a rigorous extraction of $\alpha_s$ will be already there.

For simplicity, we model the nonperturbative correction to the analytic NLO result by employing the 
``simple power correction'' ansatz from~\cite{Abdallah:2003xz} which amounts to fitting
\begin{equation}
\Sigma_H  (\chi, \alpha_s)  = \Sigma_{H,\textrm{pert}}  (\chi, \alpha_s) + C_1/m_H
\label{eq:model}
\end{equation}
to the \textsc{Pythia} data, where $\Sigma_{H,\textrm{pert}}  (\chi, \alpha_s)$ corresponds to eq.\,\eqref{eq:eecnlo} with $N_f = 5, N_c = 3$ and $\alpha_s$ being a fit parameter. Another fit parameter is $C_1$ which is assumed to be a constant that accounts for the nonperturbative corrections. Regarding the fit region, it is important to ensure that the values of $ \cos \chi $ are sufficiently far away from the collinear and back-to-back regions, where the pure fixed-order result ceases to be valid. 

\begin{figure}[ht]
\centering
\includegraphics[width=0.9\textwidth,clip]{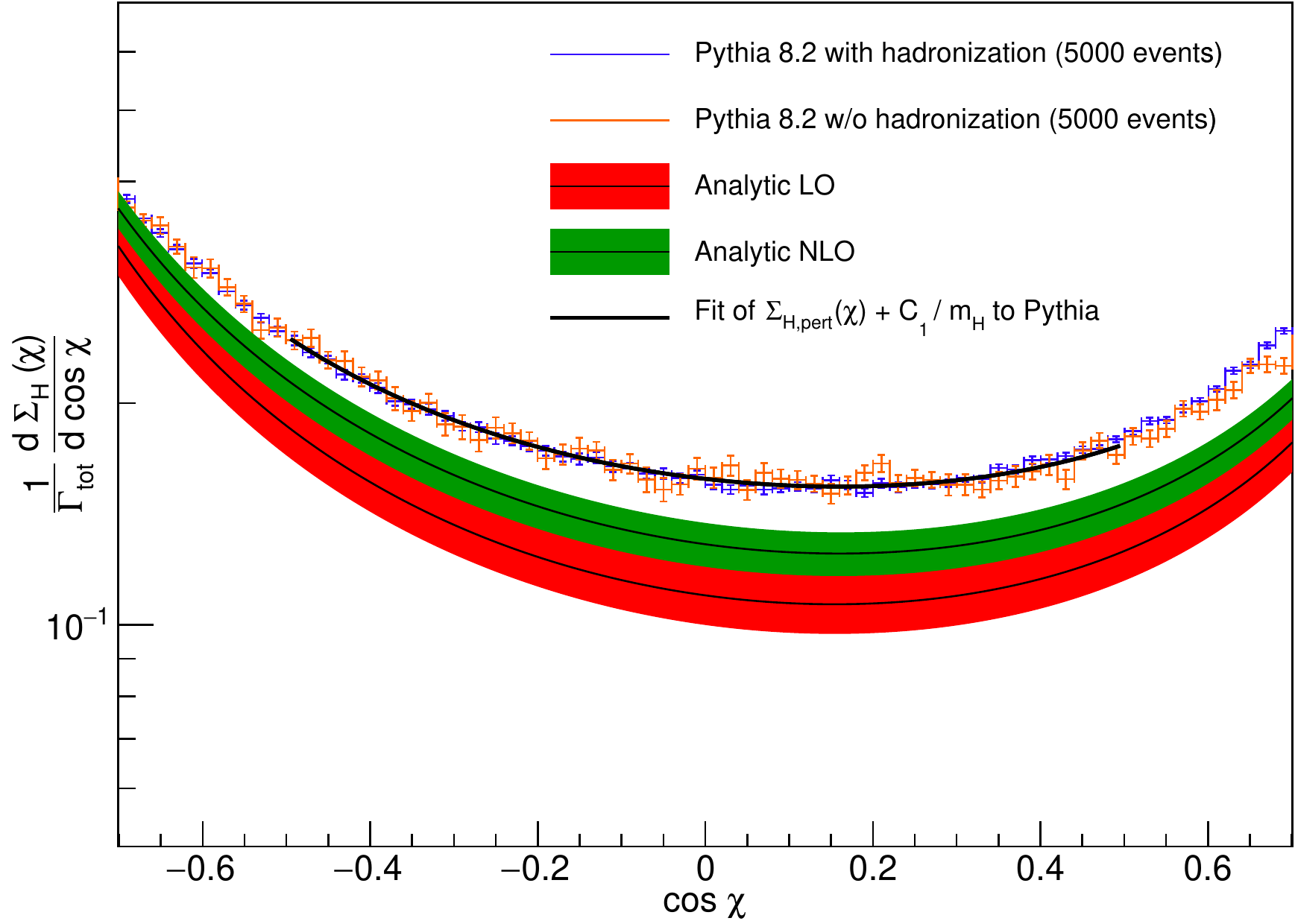}
\caption{Fit of eq.\,\ref{eq:model} to the \textsc{Pythia} data for $|\cos \chi|\leq 0.5$.}
\label{fig:pythiafit}
\end{figure}

The results of our binned maximum likelihood fit to the 5000 events simulated with \textsc{Pythia} are summarized in table\,\ref{tab:pythiafit}. Despite this being a toy fit, the fitted values of $\alpha_s$ for different  $\cos \chi$-regions are consistent with $\alpha_s(m_H) = 0.113$ within the (admittedly very large) error bounds. The fitted curve for $|\cos \chi|\leq 0.5$ is shown in figure \ref{fig:pythiafit}.

\begin{table}[ht]
\centering
\begin{tabular}{ |c  c  c  c|}
\hline 
Fit region & Fitted $\alpha_s$ & Fitted $C_1$ in GeV & \textsc{Minuit} / \textsc{Migrad} $\chi^2 / \textrm{NDF}$ \\
    \hline			
$-0.40 \leq \cos \chi \leq 0.40$ & $0.115 \pm 0.033$ & $3.36 \pm 5.56$ & $24.6/38$ \\
$-0.45 \leq \cos \chi \leq 0.45$ & $0.114 \pm 0.023$ & $3.41 \pm 3.95$ & $30.7/42$ \\
$-0.50 \leq \cos \chi \leq 0.50$ & $0.115 \pm 0.019$ & $3.28 \pm 3.26$ & $43.1/48$ \\
$-0.55 \leq \cos \chi \leq 0.50$ & $0.116 \pm 0.012$ & $3.28 \pm 2.14$ & $67.6/52$ \\

\hline
\end{tabular}
  \caption{Fit of eq.\,\ref{eq:model} to the \textsc{Pythia} data using \textsc{ROOT} with \textsc{Minuit} / \textsc{Migrad} as the minimizer.}
\label{tab:pythiafit}
\end{table}

\section{Summary}
\label{sec:summary}

In this paper we have initiated the study of the Energy-Energy Correlation in gluon-initiated Higgs decays, that can simultaneously probe our understanding of the strong and Higgs sectors. Using the techniques developed in our previous work on the standard EEC in $e^+ e^-$ annihilation~\cite{Dixon:2018qgp} and working in the framework of the Higgs EFT, we obtained the fully analytic result for Higgs EEC at NLO in fixed-order perturbation theory. This result bears strong similarities to the analytic NLO result for the standard EEC, in the sense that both observables are described by the same set of master integrals and can be written using the same basis of building block functions, made of classical polylogarithms up to weight 3.

Regarding the phenomenological relevance of the new observable, large QCD background accompanying Higgs decays at hadron colliders makes it very unlikely that the decay channel $H \to gg$ can be measured by the LHC experiments in the near future (if at all). However, such a measurement will be certainly possible in the clean environment of a future lepton collider, such as ILC, CEPC, FCC-ee or CLIC. Given sufficiently high statistics, the Higgs EEC could be also used for the determination of the strong coupling constant, as it was the case in the studies of the standard EEC conducted by the LEP experiments. 

In order to attract more interest to the Higgs EEC from the experimental and theoretical side, we provided a very rough estimate the size of the hadronization effects using \textsc{Pythia} 8. The simulated data hints that the nonperturbative corrections might be comparably small in the central region. Since the simulation is done using only the LO hard matrix element for $H \to g g$ implemented in \textsc{Pythia} and suffers from the absence of the $H \to ggg$ vertex and the error bars in our plots stem only from statistical uncertainties these results should be obviously taken with a grain of salt. Nonetheless, by comparing our analytic result to the \textsc{Pythia} prediction and using a naive model for nonperturbative effects, we also performed a toy determination of $\alpha_s$, obtaining a value that is (within the very large error bounds) compatible with the current PDG world average. The purpose of this exercise is to motivate the high energy physics community to explore the phenomenology of the Higgs EEC and other related event shape observables in greater details.

For the future, it would be very useful to obtain the NNLO correction to the Higgs EEC, at least numerically. 
This is important for the phenomenological studies, especially the determination of $\alpha_s$.
A fully analytic NNLO result is also very desirable, but as of now the complexity of the corresponding expressions poses a major obstacle to this endeavor. 
Although EEC currently appears to be the most convenient event shape variable for fixed-order analytic calculations, one may still wonder about the feasibility of analytic NLO results for other related observables such as thrust or the $C$ parameter. Once such results are obtained for parton production in $e^+ e^-$ annihilation, it would be a simple exercise to obtain the corresponding predictions also for the Higgs decays. 

\acknowledgments
We thank Peter Skands for his detailed explanation of the \textsc{Pythia} 8 options related to the hadronization and parton showers. We thank Anjie Gao for checking the results in the back-to-back limit, and Han-tian Zhang for useful discussions. The work of M.X.L., V.S., T.Z.Y., and H.X.Z. was supported in part by the National Science Foundation of China (11135006, 11275168, 11422544, 11375151, 11535002) and the Zhejiang University Fundamental Research Funds for the Central Universities (2017QNA3007). 

\appendix

\section{Subtopologies}
\label{sec:appendix0}
\subsection{LO}
Single subtopology for  $H \to g g g$
\begin{equation}
\{p_1, p_2, Q -p_1 - p_2, Q -p_2, Q -p_1\}
\end{equation}
Single subtopology for $H \to q \bar{q} g$
\begin{equation}
\{p_1, p_2, Q -p_1 - p_2,  p_1 + p_2, Q- p_1\}
\end{equation}

\subsection{NLO}
Ten (sub)topologies for $H \to g g g g$
\begin{subequations}
\begin{align}
& \{p_1, p_2, p_3, Q -p_1 - p_2 - p_3, Q -p_1 - p_2, Q -p_2, Q -p_1, p_1 + p_3, p_2 + p_3\}, \\
& \{p_1, p_2, p_3, Q -p_1 - p_2 - p_3, p_2 + p_3, p_1 + p_2 + p_3, Q -p_2 - p_3, Q -p_3, p_1 + p_2\}, \\
& \{p_1, p_2, p_3, Q -p_1 - p_2 - p_3, p_2 + p_3, p_1 + p_2 + p_3, Q -p_1 - p_3, Q -p_3, p_1 + p_2\}, \\ 
& \{p_1, p_2, p_3, Q -p_1 - p_2 - p_3, p_2 + p_3, Q -p_2 - p_3, Q -p_1 - p_3, Q -p_2, p_1 + p_3\}, \\
& \{p_1, p_2, p_3, Q -p_1 - p_2 - p_3, p_2 + p_3, Q -p_2 - p_3, Q -p_3, Q -p_1 - p_2, p_1 + p_2\}, \\
& \{p_1, p_2, p_3, Q -p_1 - p_2 - p_3, p_2 + p_3, Q -p_1 - p_3, Q -p_3, Q -p_2, p_1 + p_3\}, \\
& \{p_1, p_2, p_3, Q -p_1 - p_2 - p_3, p_1 + p_2, p_2 + p_3, Q -p_2 - p_3, Q -p_2, Q -p_1\}, \\
& \{p_1, p_2, p_3, Q -p_1 - p_2 - p_3, Q -p_1 - p_3, Q -p_3, Q -p_1, p_1 + p_2,  p_2 + p_3\}, \\
& \{p_1, p_2, p_3, Q -p_1 - p_2 - p_3, p_1 + p_2, Q -p_3, Q -p_1 - p_2, Q -p_1, p_1 + p_3\}, \\
& \{p_1, p_2, p_3, Q -p_1 - p_2 - p_3, Q -p_1 - p_3, Q -p_3, Q -p_1 - p_2, Q -p_2, p_1 + p_2\}
\end{align}
\end{subequations}
Two subtopologies for $H \to q \bar{q} g g$
\begin{subequations}
\begin{align}
& \{p_1, p_2, p_3, Q -p_1 - p_2 - p_3, p_2 + p_3, p_1 + p_2 + p_3, Q -p_2 - p_3, Q -p_3, p_1 + p_2\},  \\
& \{p_1, p_2, p_3, Q -p_1 - p_2 - p_3, p_2 + p_3, p_1 + p_2 + p_3, Q -p_1 - p_3, Q -p_3, p_1 + p_2\}
\end{align}
\end{subequations}
Single subtopology for $H \to q \bar{q} q \bar{q}$
\begin{align}
& \{p_1, p_2, p_3, Q -p_1 - p_2 - p_3, p_1 + p_2, p_2 + p_3, Q -p_2 - p_3,   Q -p_1 - p_2, p_1 + p_3\}
\end{align} 
Single subtopology for $H \to q \bar{q} q' \bar{q}'$
\begin{align}
& \{p_1, p_2, p_3, Q -p_1 - p_2 - p_3, p_1 + p_2, Q -p_1 - p_2, p_1 + p_3, p_1 - Q,   p_2 + p_3\}
 \end{align}

\section{Asymptotics of the color components}

\subsection{Collinear limit}

\begin{subequations}
\begin{align}
A_{H,\textrm{lc}}(z) & = \frac{7}{20} \frac{1}{z} + \frac{59}{120} + \mathcal{O}(z), \\ \nonumber \\
A_{H,N_f}(z) & = \frac{1}{20} \frac{1}{z} + \frac{1}{15} + \mathcal{O}(z), \\ \nonumber \\
B_{H,\textrm{lc}}(z) &= \frac{1}{z} \left[ -\frac{91}{600}  \log (z) -\frac{\zeta_3}{2}+\frac{97 \zeta _2}{60}+\frac{138427}{27000}\right]+\left[-2 \zeta _2+\frac{9221}{3150}\right] \log (z) \nonumber \\
& +\frac{13 \zeta_3}{2}  -\frac{47 \zeta_2}{15} +\frac{79860499}{10584000} + \mathcal{O}(z), \\ \nonumber \\
B_{H,\textrm{nlc}}(z) &= \frac{1}{z} \left[ \frac{143}{400} \log (z) -\frac{7 \zeta _2}{30}-\frac{50039}{27000}\right]+\left[\zeta_2-\frac{173}{140}\right] \log (z) \nonumber \\
& -3 \zeta_3 + \frac{4 \zeta _2}{3}  -\frac{1819763}{1058400}  + \mathcal{O}(z), \\ \nonumber \\ 
B_{H,\textrm{nnlc}}(z) &= \frac{1}{z} \left[\frac{7}{80}\log (z)-\frac{9}{800}\right]+\left[-\frac{\zeta _2}{2}+\frac{3163}{3600}\right] \log (z)\nonumber \\
& +\frac{3 \zeta (3)}{2} -\frac{41 \zeta _2}{60} -\frac{125143}{216000} + \mathcal{O}(z), \\ \nonumber \\
B_{H,N_f^2}(z) &= \frac{1}{z}  \left[\frac{1}{15} \log (z) -\frac{43}{120}\right]+\frac{2}{45} \log (z) -\frac{2207}{5400} + \mathcal{O}(z).
\end{align}
\end{subequations}

\subsection{Back-to-back limit}

\begin{subequations}
\begin{align}
A_{H,\textrm{lc}}(z) & =  \frac{1}{1-z} \left [ -\frac{1}{2} \log (1-z)-\frac{11}{12} \right ] - 6 \log (1-z) -\frac{77}{6} + \mathcal{O}(1-z), \\ \nonumber \\
A_{H,N_f}(z) & =  \frac{1}{3}\frac{1}{1-z} + 4 \log (1-z)+\frac{73}{6} + \mathcal{O}(1-z), \\ \nonumber \\
B_{H,\textrm{lc}}(z) &= \frac{1}{1-z} \biggl [
\frac{1}{2} \log^3(1-z)+\frac{11}{3} \log^2(1-z) + \left(\frac{3 \zeta_2}{2}+\frac{11}{8}\right) \log (1-z) \nonumber \\
& + \frac{\zeta_3}{2} + \frac{77 \zeta_2}{12}-\frac{907}{144} \biggr ]  +\frac{10}{3} \log ^3(1-z) +\frac{757}{24} \log^2(1-z) \nonumber \\
& +\left[16 \zeta_2+\frac{415}{18}\right] \log (1-z) +\frac{155 \zeta_3}{4} +15 \zeta_2 \log (2) + \frac{7001 \zeta_2}{96}-\frac{7115}{54} + \mathcal{O}(1-z), \\ \nonumber \\
B_{H,\textrm{nlc}}(z) & = \frac{1}{1-z}  \left [-\frac{4}{3} \log ^2(1-z)-\frac{17}{6} \log (1-z)-\frac{7 \zeta_2}{3}+\frac{251}{72} \right ] -2 \log ^3(1-z) \nonumber \\
&-\frac{82}{3} \log ^2(1-z)+\left[-11 \zeta_2-\frac{182}{9}\right] \log (1-z) \nonumber \\
&-27 \zeta_3-\frac{3859 \zeta_2}{48} +\frac{76291}{432} + \mathcal{O}(1-z), \\ \nonumber \\
B_{H,\textrm{nnlc}}(z) & = -\frac{1}{4 (1-z)} +\frac{1}{12} \log ^3(1-z) +\frac{7}{4} \log ^2(1-z) +\left[\frac{17}{2}-\frac{\zeta _2}{2}\right] \log (1-z)\nonumber \\
& +\frac{3 \zeta_3}{4} +3 \zeta_2 \log (2)  -\frac{151 \zeta_2}{24} +\frac{929}{48} + \mathcal{O}(1-z), \\ \nonumber \\
B_{H,N_f^2}(z) & = \frac{1}{1-z} \left [\frac{2}{3} \log (1-z)-\frac{5}{18}\right ] +4 \log^2(1-z)\nonumber \\
& +\frac{29}{9} \log (1-z) + 8 \zeta _2 -\frac{2279}{54} + \mathcal{O}(1-z).
\end{align}
\end{subequations}
\section{Identical-quark interference contributions}

In this appendix, we provide the identical-quark interference terms that correspond to the $q\bar{q}q\bar{q}$ cut diagram 
from figure \ref{fig:higgsdiags} $(\text{c})$, which contributes to $B_{H,\textrm{nnlc}}$. In addition, $B_{H,\textrm{nnlc}}$ also receives contributions from the $q\bar{q}  gg $ cut diagram from figure \ref{fig:higgsdiags} $(\text{b})$. We can write the real contributions to  $B_{H,\textrm{nnlc}}$ as
\begin{align}
 B^R_{H,\textrm{nnlc}} = B^R_{H,g} + B_{H,\textrm{qq}_{\textrm{int}}}. 
\end{align} 

Similarly with $B_{\textrm{qq}_{\textrm{int}}}$ from the standard EEC~\citep{Dixon:2018qgp}, there are no virtual corrections to $B_{H,\textrm{qq}_{\textrm{int}}}$ at NLO. Thus, this contribution is separately IR finite and gauge invariant and can be written as 
\begin{align}
B_{H,\textrm{qq}_{\textrm{int}}} &= -\frac{483 z^2-1075 z-8}{72 z^5} +\frac{110 z^3-729 z^2+1116 z-326}{36 z^6}  g_1^{(1)}  \nonumber\\
&-\frac{47 z^2-231 z+330}{36 z^5}  g_2^{(1)} -\frac{19 z^3-63 z^2+87 z-55 }{12 z^6} g_1^{(2)} \nonumber\\
&-\frac{2 z^3-21 z^2+63 z-55 }{6 z^6} g_2^{(2)}+\frac{21 z^3-84 z^2+150 z-110 }{6 z^6} g_4^{(2)} \nonumber\\
&-\frac{4 z^3-25 z^2+38 z-18 }{24 z^6} g_2^{(3)}+ \frac{z^2-2 z+2 }{8 z^6} \left(g_6^{(3)}-2 g_7^{(3)}\right),
\end{align}
where we introduced two additional building block functions of pure transcendental weight 3, in addition to those already presented in eq.\,\eqref{eq:gdef}
\begin{align}
 g_{6}^{(3)} = \; & \log ^3(1-z)-15 \zeta_2 \log (1-z) \,,
\nonumber\\
  g_{7}^{(3)} = \; & \log (1-z) \left(\text{Li}_2(z)+\log (1-z) \log (z)-\frac{15 \zeta_2}{2}\right) \,.
\end{align} 
It is interesting to observe that these two functions always appear together as $(g_6^{(3)}-2 g_7^{(3)})$, which is also true for $B_{\textrm{qq}_{\textrm{int}}}$ of the standard EEC~\citep{Dixon:2018qgp}. However, $B_{H,\textrm{qq}_{\textrm{int}}}$ is much simpler than $B_{\textrm{qq}_{\textrm{int}}}$. While the latter contains 11 building block functions, including those with an explicit dependence on $\sqrt{z}$ and $i \sqrt{z}/\sqrt{1-z}$, the former can be expressed in terms of only 8 functions, all of which depend solely on $z$.

Expanding the result for $B_{H,\textrm{qq}_{\textrm{int}}}$ in the collinear limit we get 
\begin{align}
B_{H,\textrm{qq}_{\textrm{int}}}(z) = -\frac{\log (z)}{720}-\frac{101}{14400} + \mathcal{O}(z),
\end{align}
while the expansion the back-to-back limit yields 
\begin{align}
B_{H,\textrm{qq}_{\textrm{int}}} & = \left(\frac{19}{4}-\frac{\zeta _2}{2}\right) \log (1-z)+\frac{1}{12} \log ^3(1-z)  \nonumber \\
& +\log ^2(1-z)-\frac{5 \zeta _2}{3}-\frac{\zeta
   _3}{2}+\frac{25}{3} \mathcal{O}(1-z). 
\end{align} 
We would like to stress the fact that both limits are free of leading-power singularities $1/z$ and $1/(1-z)$ respectively. This is mainly because quark-antiquark pairs can only arise in the splitting of gluons originating directly from the decay of the Higgs and those gluons cannot be soft. 

\bibliographystyle{JHEP}

\begin{thebibliography}{100}

\bibitem{CEPCStudyGroup:2018rmc}
{\scshape CEPC Study Group} collaboration, \emph{{CEPC Conceptual Design
  Report: Volume 1 - Accelerator}},
  \href{https://arxiv.org/abs/1809.00285}{{\ttfamily 1809.00285}}.

\bibitem{CEPCStudyGroup:2018ghi}
{\scshape CEPC Study Group} collaboration, \emph{{CEPC Conceptual Design
  Report: Volume 2 - Physics \& Detector}},
  \href{https://arxiv.org/abs/1811.10545}{{\ttfamily 1811.10545}}.

\bibitem{Behnke:2013xla}
T.~Behnke, J.~E. Brau, B.~Foster, J.~Fuster, M.~Harrison, J.~M. Paterson
  et~al., \emph{{The International Linear Collider Technical Design Report -
  Volume 1: Executive Summary}},
  \href{https://arxiv.org/abs/1306.6327}{{\ttfamily 1306.6327}}.

\bibitem{Baer:2013cma}
H.~Baer, T.~Barklow, K.~Fujii, Y.~Gao, A.~Hoang, S.~Kanemura et~al., \emph{{The
  International Linear Collider Technical Design Report - Volume 2: Physics}},
  \href{https://arxiv.org/abs/1306.6352}{{\ttfamily 1306.6352}}.

\bibitem{Gomez-Ceballos:2013zzn}
{\scshape TLEP Design Study Working Group} collaboration, \emph{{First Look at
  the Physics Case of TLEP}},
  \href{https://doi.org/10.1007/JHEP01(2014)164}{\emph{JHEP} {\bfseries 01}
  (2014) 164} [\href{https://arxiv.org/abs/1308.6176}{{\ttfamily 1308.6176}}].

\bibitem{Aicheler:2012bya}
M.~Aicheler, P.~Burrows, M.~Draper, T.~Garvey, P.~Lebrun, K.~Peach et~al.,
  \emph{{A Multi-TeV Linear Collider Based on CLIC Technology}}, .

\bibitem{deBlas:2018mhx}
J.~de~Blas et~al., \emph{{The CLIC Potential for New Physics}},
  \href{https://arxiv.org/abs/1812.02093}{{\ttfamily 1812.02093}}.

\bibitem{Heister:2003aj}
{\scshape ALEPH} collaboration, \emph{{Studies of QCD at e+ e- centre-of-mass
  energies between 91-GeV and 209-GeV}},
  \href{https://doi.org/10.1140/epjc/s2004-01891-4}{\emph{Eur. Phys. J.}
  {\bfseries C35} (2004) 457}.

\bibitem{Abdallah:2004xe}
{\scshape DELPHI} collaboration, \emph{{The Measurement of alpha(s) from event
  shapes with the DELPHI detector at the highest LEP energies}},
  \href{https://doi.org/10.1140/epjc/s2004-01889-x}{\emph{Eur. Phys. J.}
  {\bfseries C37} (2004) 1}
  [\href{https://arxiv.org/abs/hep-ex/0406011}{{\ttfamily hep-ex/0406011}}].

\bibitem{Achard:2004sv}
{\scshape L3} collaboration, \emph{{Studies of hadronic event structure in
  $e^{+} e^{-}$ annihilation from 30-GeV to 209-GeV with the L3 detector}},
  \href{https://doi.org/10.1016/j.physrep.2004.07.002}{\emph{Phys. Rept.}
  {\bfseries 399} (2004) 71}
  [\href{https://arxiv.org/abs/hep-ex/0406049}{{\ttfamily hep-ex/0406049}}].

\bibitem{Abbiendi:2004qz}
{\scshape OPAL} collaboration, \emph{{Measurement of event shape distributions
  and moments in e+ e- ---> hadrons at 91-GeV - 209-GeV and a determination of
  alpha(s)}}, \href{https://doi.org/10.1140/epjc/s2005-02120-6}{\emph{Eur.
  Phys. J.} {\bfseries C40} (2005) 287}
  [\href{https://arxiv.org/abs/hep-ex/0503051}{{\ttfamily hep-ex/0503051}}].

\bibitem{Brandt:1964sa}
S.~Brandt, C.~Peyrou, R.~Sosnowski and A.~Wroblewski, \emph{{The Principal axis
  of jets. An Attempt to analyze high-energy collisions as two-body
  processes}}, \href{https://doi.org/10.1016/0031-9163(64)91176-X}{\emph{Phys.
  Lett.} {\bfseries 12} (1964) 57}.

\bibitem{Farhi:1977sg}
E.~Farhi, \emph{{A QCD Test for Jets}},
  \href{https://doi.org/10.1103/PhysRevLett.39.1587}{\emph{Phys. Rev. Lett.}
  {\bfseries 39} (1977) 1587}.

\bibitem{Clavelli:1981yh}
L.~Clavelli and D.~Wyler, \emph{{Kinematical Bounds on Jet Variables and the
  Heavy Jet Mass Distribution}},
  \href{https://doi.org/10.1016/0370-2693(81)90248-3}{\emph{Phys. Lett.}
  {\bfseries 103B} (1981) 383}.

\bibitem{Rakow:1981qn}
P.~E.~L. Rakow and B.~R. Webber, \emph{{Transverse Momentum Moments of Hadron
  Distributions in {QCD} Jets}},
  \href{https://doi.org/10.1016/0550-3213(81)90286-8}{\emph{Nucl. Phys.}
  {\bfseries B191} (1981) 63}.

\bibitem{Ellis:1986ig}
R.~K. Ellis and B.~R. Webber, \emph{{QCD Jet Broadening in Hadron Hadron
  Collisions}}, {\emph{Conf. Proc.} {\bfseries C860623} (1986) 74}.

\bibitem{Catani:1992jc}
S.~Catani, G.~Turnock and B.~R. Webber, \emph{{Jet broadening measures in
  $e^{+} e^{-}$ annihilation}},
  \href{https://doi.org/10.1016/0370-2693(92)91565-Q}{\emph{Phys. Lett.}
  {\bfseries B295} (1992) 269}.

\bibitem{Parisi:1978eg}
G.~Parisi, \emph{{Super Inclusive Cross-Sections}},
  \href{https://doi.org/10.1016/0370-2693(78)90061-8}{\emph{Phys. Lett.}
  {\bfseries 74B} (1978) 65}.

\bibitem{Donoghue:1979vi}
J.~F. Donoghue, F.~E. Low and S.-Y. Pi, \emph{{Tensor Analysis of Hadronic Jets
  in Quantum Chromodynamics}},
  \href{https://doi.org/10.1103/PhysRevD.20.2759}{\emph{Phys. Rev.} {\bfseries
  D20} (1979) 2759}.

\bibitem{Catani:1991hj}
S.~Catani, Y.~L. Dokshitzer, M.~Olsson, G.~Turnock and B.~R. Webber, \emph{{New
  clustering algorithm for multi - jet cross-sections in e+ e- annihilation}},
  \href{https://doi.org/10.1016/0370-2693(91)90196-W}{\emph{Phys. Lett.}
  {\bfseries B269} (1991) 432}.

\bibitem{Aktas:2005tz}
{\scshape H1} collaboration, \emph{{Measurement of event shape variables in
  deep-inelastic scattering at HERA}},
  \href{https://doi.org/10.1140/epjc/s2006-02493-x}{\emph{Eur. Phys. J.}
  {\bfseries C46} (2006) 343}
  [\href{https://arxiv.org/abs/hep-ex/0512014}{{\ttfamily hep-ex/0512014}}].

\bibitem{Aaltonen:2011et}
{\scshape CDF} collaboration, \emph{{Measurement of Event Shapes in
  Proton-Antiproton Collisions at Center-of-Mass Energy 1.96 TeV}},
  \href{https://doi.org/10.1103/PhysRevD.83.112007}{\emph{Phys. Rev.}
  {\bfseries D83} (2011) 112007}
  [\href{https://arxiv.org/abs/1103.5143}{{\ttfamily 1103.5143}}].

\bibitem{Banfi:2010xy}
A.~Banfi, G.~P. Salam and G.~Zanderighi, \emph{{Phenomenology of event shapes
  at hadron colliders}},
  \href{https://doi.org/10.1007/JHEP06(2010)038}{\emph{JHEP} {\bfseries 06}
  (2010) 038} [\href{https://arxiv.org/abs/1001.4082}{{\ttfamily 1001.4082}}].

\bibitem{Gehrmann-DeRidder:2007nzq}
A.~Gehrmann-De~Ridder, T.~Gehrmann, E.~W.~N. Glover and G.~Heinrich,
  \emph{{Second-order QCD corrections to the thrust distribution}},
  \href{https://doi.org/10.1103/PhysRevLett.99.132002}{\emph{Phys. Rev. Lett.}
  {\bfseries 99} (2007) 132002}
  [\href{https://arxiv.org/abs/0707.1285}{{\ttfamily 0707.1285}}].

\bibitem{GehrmannDeRidder:2007hr}
A.~Gehrmann-De~Ridder, T.~Gehrmann, E.~W.~N. Glover and G.~Heinrich,
  \emph{{NNLO corrections to event shapes in e+ e- annihilation}},
  \href{https://doi.org/10.1088/1126-6708/2007/12/094}{\emph{JHEP} {\bfseries
  12} (2007) 094} [\href{https://arxiv.org/abs/0711.4711}{{\ttfamily
  0711.4711}}].

\bibitem{Weinzierl:2009ms}
S.~Weinzierl, \emph{{Event shapes and jet rates in electron-positron
  annihilation at NNLO}},
  \href{https://doi.org/10.1088/1126-6708/2009/06/041}{\emph{JHEP} {\bfseries
  06} (2009) 041} [\href{https://arxiv.org/abs/0904.1077}{{\ttfamily
  0904.1077}}].

\bibitem{Ridder:2014wza}
A.~Gehrmann-De~Ridder, T.~Gehrmann, E.~W.~N. Glover and G.~Heinrich,
  \emph{{EERAD3: Event shapes and jet rates in electron-positron annihilation
  at order $\alpha_s^3$}},
  \href{https://doi.org/10.1016/j.cpc.2014.07.024}{\emph{Comput. Phys. Commun.}
  {\bfseries 185} (2014) 3331}
  [\href{https://arxiv.org/abs/1402.4140}{{\ttfamily 1402.4140}}].

\bibitem{DelDuca:2016ily}
V.~Del~Duca, C.~Duhr, A.~Kardos, G.~Somogyi, Z.~Szőr, Z.~Trócsányi et~al.,
  \emph{{Jet production in the CoLoRFulNNLO method: event shapes in
  electron-positron collisions}},
  \href{https://doi.org/10.1103/PhysRevD.94.074019}{\emph{Phys. Rev.}
  {\bfseries D94} (2016) 074019}
  [\href{https://arxiv.org/abs/1606.03453}{{\ttfamily 1606.03453}}].

\bibitem{deFlorian:2004mp}
D.~de~Florian and M.~Grazzini, \emph{{The Back-to-back region in e+ e-
  energy-energy correlation}},
  \href{https://doi.org/10.1016/j.nuclphysb.2004.10.051}{\emph{Nucl. Phys.}
  {\bfseries B704} (2005) 387}
  [\href{https://arxiv.org/abs/hep-ph/0407241}{{\ttfamily hep-ph/0407241}}].

\bibitem{Becher:2008cf}
T.~Becher and M.~D. Schwartz, \emph{{A precise determination of $\alpha_s$ from
  LEP thrust data using effective field theory}},
  \href{https://doi.org/10.1088/1126-6708/2008/07/034}{\emph{JHEP} {\bfseries
  07} (2008) 034} [\href{https://arxiv.org/abs/0803.0342}{{\ttfamily
  0803.0342}}].

\bibitem{Chien:2010kc}
Y.-T. Chien and M.~D. Schwartz, \emph{{Resummation of heavy jet mass and
  comparison to LEP data}},
  \href{https://doi.org/10.1007/JHEP08(2010)058}{\emph{JHEP} {\bfseries 08}
  (2010) 058} [\href{https://arxiv.org/abs/1005.1644}{{\ttfamily 1005.1644}}].

\bibitem{Abbate:2010xh}
R.~Abbate, M.~Fickinger, A.~H. Hoang, V.~Mateu and I.~W. Stewart, \emph{{Thrust
  at N$^3$LL with Power Corrections and a Precision Global Fit for
  alphas(mZ)}}, \href{https://doi.org/10.1103/PhysRevD.83.074021}{\emph{Phys.
  Rev.} {\bfseries D83} (2011) 074021}
  [\href{https://arxiv.org/abs/1006.3080}{{\ttfamily 1006.3080}}].

\bibitem{Monni:2011gb}
P.~F. Monni, T.~Gehrmann and G.~Luisoni, \emph{{Two-Loop Soft Corrections and
  Resummation of the Thrust Distribution in the Dijet Region}},
  \href{https://doi.org/10.1007/JHEP08(2011)010}{\emph{JHEP} {\bfseries 08}
  (2011) 010} [\href{https://arxiv.org/abs/1105.4560}{{\ttfamily 1105.4560}}].

\bibitem{Becher:2012qc}
T.~Becher and G.~Bell, \emph{{NNLL Resummation for Jet Broadening}},
  \href{https://doi.org/10.1007/JHEP11(2012)126}{\emph{JHEP} {\bfseries 11}
  (2012) 126} [\href{https://arxiv.org/abs/1210.0580}{{\ttfamily 1210.0580}}].

\bibitem{Banfi:2014sua}
A.~Banfi, H.~McAslan, P.~F. Monni and G.~Zanderighi, \emph{{A general method
  for the resummation of event-shape distributions in $e^{+} e^{-}$
  annihilation}}, \href{https://doi.org/10.1007/JHEP05(2015)102}{\emph{JHEP}
  {\bfseries 05} (2015) 102} [\href{https://arxiv.org/abs/1412.2126}{{\ttfamily
  1412.2126}}].

\emph{\bibitem{Hoang:2014wka}
A.~H. Hoang, D.~W. Kolodrubetz, V.~Mateu and I.~W. Stewart,
  \emph{{$C$-parameter distribution at N$^3$LL' including power
  corrections}}, \href{https://doi.org/10.1103/PhysRevD.91.094017}{\emph{Phys.
  Rev.} {\bfseries D91} (2015) 094017}
  [\href{https://arxiv.org/abs/1411.6633}{{\ttfamily 1411.6633}}].}

\bibitem{Banfi:2016zlc}
A.~Banfi, H.~McAslan, P.~F. Monni and G.~Zanderighi, \emph{{The two-jet rate in
  $e^+e^-$ at next-to-next-to-leading-logarithmic order}},
  \href{https://doi.org/10.1103/PhysRevLett.117.172001}{\emph{Phys. Rev. Lett.}
  {\bfseries 117} (2016) 172001}
  [\href{https://arxiv.org/abs/1607.03111}{{\ttfamily 1607.03111}}].

\bibitem{Tulipant:2017ybb}
Z.~Tulipant, A.~Kardos and G.~Somogyi, \emph{{Energy–energy correlation in
  electron–positron annihilation at NNLL + NNLO accuracy}},
  \href{https://doi.org/10.1140/epjc/s10052-017-5320-9}{\emph{Eur. Phys. J.}
  {\bfseries C77} (2017) 749}
  [\href{https://arxiv.org/abs/1708.04093}{{\ttfamily 1708.04093}}].

\bibitem{Moult:2018jzp}
I.~Moult and H.~X. Zhu, \emph{{Simplicity from Recoil: The Three-Loop Soft
  Function and Factorization for the Energy-Energy Correlation}},
  \href{https://doi.org/10.1007/JHEP08(2018)160}{\emph{JHEP} {\bfseries 08}
  (2018) 160} [\href{https://arxiv.org/abs/1801.02627}{{\ttfamily
  1801.02627}}].

\bibitem{Kardos:2018kqj}
A.~Kardos, S.~Kluth, G.~Somogyi, Z.~Tulipant and A.~Verbytskyi, \emph{{Precise
  determination of $\alpha _{S}(M_Z)$ from a global fit of energy–energy
  correlation to NNLO+NNLL predictions}},
  \href{https://doi.org/10.1140/epjc/s10052-018-5963-1}{\emph{Eur. Phys. J.}
  {\bfseries C78} (2018) 498}
  [\href{https://arxiv.org/abs/1804.09146}{{\ttfamily 1804.09146}}].

\bibitem{Banfi:2018mcq}
A.~Banfi, B.~K. El-Menoufi and P.~F. Monni, \emph{{The Sudakov radiator for jet
  observables and the soft physical coupling}},
  \href{https://doi.org/10.1007/JHEP01(2019)083}{\emph{JHEP} {\bfseries 01}
  (2019) 083} [\href{https://arxiv.org/abs/1807.11487}{{\ttfamily
  1807.11487}}].

\bibitem{Bell:2018gce}
G.~Bell, A.~Hornig, C.~Lee and J.~Talbert, \emph{{$e^+ e^-$ angularity
  distributions at NNLL$^\prime$ accuracy}},
  \href{https://doi.org/10.1007/JHEP01(2019)147}{\emph{JHEP} {\bfseries 01}
  (2019) 147} [\href{https://arxiv.org/abs/1808.07867}{{\ttfamily
  1808.07867}}].

\bibitem{Verbytskyi:2019zhh}
A.~Verbytskyi, A.~Banfi, A.~Kardos, P.~F. Monni, S.~Kluth, G.~Somogyi et~al.,
  \emph{{High precision determination of $\alpha_s$ from a global fit of jet
  rates}}, {\emph{Submitted to: JHEP} (2019) }
  [\href{https://arxiv.org/abs/1902.08158}{{\ttfamily 1902.08158}}].

\bibitem{Nagy:2001fj}
Z.~Nagy, \emph{{Three jet cross-sections in hadron hadron collisions at
  next-to-leading order}},
  \href{https://doi.org/10.1103/PhysRevLett.88.122003}{\emph{Phys. Rev. Lett.}
  {\bfseries 88} (2002) 122003}
  [\href{https://arxiv.org/abs/hep-ph/0110315}{{\ttfamily hep-ph/0110315}}].

\bibitem{Nagy:2003tz}
Z.~Nagy, \emph{{Next-to-leading order calculation of three jet observables in
  hadron hadron collision}},
  \href{https://doi.org/10.1103/PhysRevD.68.094002}{\emph{Phys. Rev.}
  {\bfseries D68} (2003) 094002}
  [\href{https://arxiv.org/abs/hep-ph/0307268}{{\ttfamily hep-ph/0307268}}].

\bibitem{Catani:1996jh}
S.~Catani and M.~H. Seymour, \emph{{The Dipole formalism for the calculation of
  QCD jet cross-sections at next-to-leading order}},
  \href{https://doi.org/10.1016/0370-2693(96)00425-X}{\emph{Phys. Lett.}
  {\bfseries B378} (1996) 287}
  [\href{https://arxiv.org/abs/hep-ph/9602277}{{\ttfamily hep-ph/9602277}}].

\bibitem{Catani:1996vz}
S.~Catani and M.~H. Seymour, \emph{{A General algorithm for calculating jet
  cross-sections in NLO QCD}},
  \href{https://doi.org/10.1016/S0550-3213(96)00589-5,
  10.1016/S0550-3213(98)81022-5}{\emph{Nucl. Phys.} {\bfseries B485} (1997)
  291} [\href{https://arxiv.org/abs/hep-ph/9605323}{{\ttfamily
  hep-ph/9605323}}].

\bibitem{Gao:2016jcm}
J.~Gao, \emph{{Probing light-quark Yukawa couplings via hadronic event shapes
  at lepton colliders}},
  \href{https://doi.org/10.1007/JHEP01(2018)038}{\emph{JHEP} {\bfseries 01}
  (2018) 038} [\href{https://arxiv.org/abs/1608.01746}{{\ttfamily
  1608.01746}}].

\bibitem{Basham:1978bw}
C.~L. Basham, L.~S. Brown, S.~D. Ellis and S.~T. Love, \emph{{Energy
  Correlations in electron - Positron Annihilation: Testing QCD}},
  \href{https://doi.org/10.1103/PhysRevLett.41.1585}{\emph{Phys. Rev. Lett.}
  {\bfseries 41} (1978) 1585}.

\bibitem{Wilczek:1977zn}
F.~Wilczek, \emph{{Decays of Heavy Vector Mesons Into Higgs Particles}},
  \href{https://doi.org/10.1103/PhysRevLett.39.1304}{\emph{Phys. Rev. Lett.}
  {\bfseries 39} (1977) 1304}.

\bibitem{Shifman:1978zn}
M.~A. Shifman, A.~I. Vainshtein and V.~I. Zakharov, \emph{{Remarks on Higgs
  Boson Interactions with Nucleons}},
  \href{https://doi.org/10.1016/0370-2693(78)90481-1}{\emph{Phys. Lett.}
  {\bfseries 78B} (1978) 443}.

\bibitem{Inami:1982xt}
T.~Inami, T.~Kubota and Y.~Okada, \emph{{Effective Gauge Theory and the Effect
  of Heavy Quarks in Higgs Boson Decays}},
  \href{https://doi.org/10.1007/BF01571710}{\emph{Z. Phys.} {\bfseries C18}
  (1983) 69}.

\bibitem{Kniehl:1995tn}
B.~A. Kniehl and M.~Spira, \emph{{Low-energy theorems in Higgs physics}},
  \href{https://doi.org/10.1007/s002880050007}{\emph{Z. Phys.} {\bfseries C69}
  (1995) 77} [\href{https://arxiv.org/abs/hep-ph/9505225}{{\ttfamily
  hep-ph/9505225}}].

\bibitem{Baikov:2016tgj}
P.~A. Baikov, K.~G. Chetyrkin and J.~H. Kühn, \emph{{Five-Loop Running of the
  QCD coupling constant}},
  \href{https://doi.org/10.1103/PhysRevLett.118.082002}{\emph{Phys. Rev. Lett.}
  {\bfseries 118} (2017) 082002}
  [\href{https://arxiv.org/abs/1606.08659}{{\ttfamily 1606.08659}}].

\bibitem{Chetyrkin:1997iv}
K.~G. Chetyrkin, B.~A. Kniehl and M.~Steinhauser, \emph{{Hadronic Higgs decay
  to order alpha-s**4}},
  \href{https://doi.org/10.1103/PhysRevLett.79.353}{\emph{Phys. Rev. Lett.}
  {\bfseries 79} (1997) 353}
  [\href{https://arxiv.org/abs/hep-ph/9705240}{{\ttfamily hep-ph/9705240}}].

\bibitem{Dixon:2018qgp}
L.~J. Dixon, M.-X. Luo, V.~Shtabovenko, T.-Z. Yang and H.~X. Zhu,
  \emph{{Analytical Computation of Energy-Energy Correlation at Next-to-Leading
  Order in QCD}},
  \href{https://doi.org/10.1103/PhysRevLett.120.102001}{\emph{Phys. Rev. Lett.}
  {\bfseries 120} (2018) 102001}
  [\href{https://arxiv.org/abs/1801.03219}{{\ttfamily 1801.03219}}].

\bibitem{Belitsky:2013xxa}
A.~V. Belitsky, S.~Hohenegger, G.~P. Korchemsky, E.~Sokatchev and A.~Zhiboedov,
  \emph{{From correlation functions to event shapes}},
  \href{https://doi.org/10.1016/j.nuclphysb.2014.04.020}{\emph{Nucl. Phys.}
  {\bfseries B884} (2014) 305}
  [\href{https://arxiv.org/abs/1309.0769}{{\ttfamily 1309.0769}}].

\bibitem{Belitsky:2013bja}
A.~V. Belitsky, S.~Hohenegger, G.~P. Korchemsky, E.~Sokatchev and A.~Zhiboedov,
  \emph{{Event shapes in $\mathcal{N} = 4$ super-Yang-Mills theory}},
  \href{https://doi.org/10.1016/j.nuclphysb.2014.04.019}{\emph{Nucl. Phys.}
  {\bfseries B884} (2014) 206}
  [\href{https://arxiv.org/abs/1309.1424}{{\ttfamily 1309.1424}}].

\bibitem{Belitsky:2013ofa}
A.~V. Belitsky, S.~Hohenegger, G.~P. Korchemsky, E.~Sokatchev and A.~Zhiboedov,
  \emph{{Energy-Energy Correlations in N=4 Supersymmetric Yang-Mills Theory}},
  \href{https://doi.org/10.1103/PhysRevLett.112.071601}{\emph{Phys. Rev. Lett.}
  {\bfseries 112} (2014) 071601}
  [\href{https://arxiv.org/abs/1311.6800}{{\ttfamily 1311.6800}}].

\bibitem{Henn:2019gkr}
J.~M. Henn, E.~Sokatchev, K.~Yan and A.~Zhiboedov, \emph{{Energy-energy
  correlations at next-to-next-to-leading order}},
  \href{https://arxiv.org/abs/1903.05314}{{\ttfamily 1903.05314}}.

\bibitem{Gao:2019mlt}
J.~Gao, Y.~Gong, W.-L. Ju and L.~L. Yang, \emph{{Thrust distribution in Higgs
  decays at the next-to-leading order and beyond}},
  \href{https://arxiv.org/abs/1901.02253}{{\ttfamily 1901.02253}}.

\bibitem{Anastasiou:2002yz}
C.~Anastasiou and K.~Melnikov, \emph{{Higgs boson production at hadron
  colliders in NNLO QCD}},
  \href{https://doi.org/10.1016/S0550-3213(02)00837-4}{\emph{Nucl. Phys.}
  {\bfseries B646} (2002) 220}
  [\href{https://arxiv.org/abs/hep-ph/0207004}{{\ttfamily hep-ph/0207004}}].

\bibitem{Anastasiou:2003yy}
C.~Anastasiou, L.~J. Dixon, K.~Melnikov and F.~Petriello, \emph{{Dilepton
  rapidity distribution in the Drell-Yan process at NNLO in QCD}},
  \href{https://doi.org/10.1103/PhysRevLett.91.182002}{\emph{Phys. Rev. Lett.}
  {\bfseries 91} (2003) 182002}
  [\href{https://arxiv.org/abs/hep-ph/0306192}{{\ttfamily hep-ph/0306192}}].

\bibitem{Chetyrkin:1981qh}
K.~G. Chetyrkin and F.~V. Tkachov, \emph{{Integration by Parts: The Algorithm
  to Calculate beta Functions in 4 Loops}},
  \href{https://doi.org/10.1016/0550-3213(81)90199-1}{\emph{Nucl. Phys.}
  {\bfseries B192} (1981) 159}.

\bibitem{Tkachov:1981wb}
F.~V. Tkachov, \emph{{A Theorem on Analytical Calculability of Four Loop
  Renormalization Group Functions}},
  \href{https://doi.org/10.1016/0370-2693(81)90288-4}{\emph{Phys. Lett.}
  {\bfseries 100B} (1981) 65}.

\bibitem{Kotikov:1991pm}
A.~V. Kotikov, \emph{{Differential equation method: The Calculation of N point
  Feynman diagrams}}, \href{https://doi.org/10.1016/0370-2693(91)90536-Y,
  10.1016/0370-2693(92)91582-T}{\emph{Phys. Lett.} {\bfseries B267} (1991)
  123}.

\bibitem{Kotikov:1990kg}
A.~V. Kotikov, \emph{{Differential equations method: New technique for massive
  Feynman diagrams calculation}},
  \href{https://doi.org/10.1016/0370-2693(91)90413-K}{\emph{Phys. Lett.}
  {\bfseries B254} (1991) 158}.

\bibitem{Kotikov:1991hm}
A.~V. Kotikov, \emph{{Differential equations method: The Calculation of vertex
  type Feynman diagrams}},
  \href{https://doi.org/10.1016/0370-2693(91)90834-D}{\emph{Phys. Lett.}
  {\bfseries B259} (1991) 314}.

\bibitem{Bern:1993kr}
Z.~Bern, L.~J. Dixon and D.~A. Kosower, \emph{{Dimensionally regulated pentagon
  integrals}}, \href{https://doi.org/10.1016/0550-3213(94)90398-0}{\emph{Nucl.
  Phys.} {\bfseries B412} (1994) 751}
  [\href{https://arxiv.org/abs/hep-ph/9306240}{{\ttfamily hep-ph/9306240}}].

\bibitem{Remiddi:1997ny}
E.~Remiddi, \emph{{Differential equations for Feynman graph amplitudes}},
  {\emph{Nuovo Cim.} {\bfseries A110} (1997) 1435}
  [\href{https://arxiv.org/abs/hep-th/9711188}{{\ttfamily hep-th/9711188}}].

\bibitem{Gehrmann:1999as}
T.~Gehrmann and E.~Remiddi, \emph{{Differential equations for two loop four
  point functions}},
  \href{https://doi.org/10.1016/S0550-3213(00)00223-6}{\emph{Nucl. Phys.}
  {\bfseries B580} (2000) 485}
  [\href{https://arxiv.org/abs/hep-ph/9912329}{{\ttfamily hep-ph/9912329}}].

\bibitem{Henn:2013pwa}
J.~M. Henn, \emph{{Multiloop integrals in dimensional regularization made
  simple}}, \href{https://doi.org/10.1103/PhysRevLett.110.251601}{\emph{Phys.
  Rev. Lett.} {\bfseries 110} (2013) 251601}
  [\href{https://arxiv.org/abs/1304.1806}{{\ttfamily 1304.1806}}].

\bibitem{Gituliar:2017umx}
O.~Gituliar and S.~Moch, \emph{{Fuchsia and Master Integrals for Energy-Energy
  Correlations at NLO in QCD}}, \href{https://doi.org/10.22323/1.290.0038,
  10.5506/APhysPolB.48.2355}{\emph{Acta Phys. Polon.} {\bfseries B48} (2017)
  2355} [\href{https://arxiv.org/abs/1711.05549}{{\ttfamily 1711.05549}}].

\bibitem{Nogueira:1991ex}
P.~Nogueira, \emph{{Automatic Feynman graph generation}},
  \href{https://doi.org/10.1006/jcph.1993.1074}{\emph{J. Comput. Phys.}
  {\bfseries 105} (1993) 279}.

\bibitem{Hahn:2000kx}
T.~Hahn, \emph{{Generating Feynman diagrams and amplitudes with FeynArts 3}},
  \href{https://doi.org/10.1016/S0010-4655(01)00290-9}{\emph{Comput. Phys.
  Commun.} {\bfseries 140} (2001) 418}
  [\href{https://arxiv.org/abs/hep-ph/0012260}{{\ttfamily hep-ph/0012260}}].

\bibitem{Alloul:2013bka}
A.~Alloul, N.~D. Christensen, C.~Degrande, C.~Duhr and B.~Fuks,
  \emph{{FeynRules 2.0 - A complete toolbox for tree-level phenomenology}},
  \href{https://doi.org/10.1016/j.cpc.2014.04.012}{\emph{Comput. Phys. Commun.}
  {\bfseries 185} (2014) 2250}
  [\href{https://arxiv.org/abs/1310.1921}{{\ttfamily 1310.1921}}].

\bibitem{Mertig:1990an}
R.~Mertig, M.~Bohm and A.~Denner, \emph{{FEYN CALC: Computer algebraic
  calculation of Feynman amplitudes}},
  \href{https://doi.org/10.1016/0010-4655(91)90130-D}{\emph{Comput. Phys.
  Commun.} {\bfseries 64} (1991) 345}.

\bibitem{Shtabovenko:2016sxi}
V.~Shtabovenko, R.~Mertig and F.~Orellana, \emph{{New Developments in FeynCalc
  9.0}}, \href{https://doi.org/10.1016/j.cpc.2016.06.008}{\emph{Comput. Phys.
  Commun.} {\bfseries 207} (2016) 432}
  [\href{https://arxiv.org/abs/1601.01167}{{\ttfamily 1601.01167}}].

\bibitem{Vermaseren:2000nd}
J.~A.~M. Vermaseren, \emph{{New features of FORM}},
  \href{https://arxiv.org/abs/math-ph/0010025}{{\ttfamily math-ph/0010025}}.

\bibitem{vanRitbergen:1998pn}
T.~van Ritbergen, A.~N. Schellekens and J.~A.~M. Vermaseren, \emph{{Group
  theory factors for Feynman diagrams}},
  \href{https://doi.org/10.1142/S0217751X99000038}{\emph{Int. J. Mod. Phys.}
  {\bfseries A14} (1999) 41}
  [\href{https://arxiv.org/abs/hep-ph/9802376}{{\ttfamily hep-ph/9802376}}].

\bibitem{Feng:2012iq}
F.~Feng, \emph{{\$Apart: A Generalized Mathematica Apart Function}},
  \href{https://doi.org/10.1016/j.cpc.2012.03.025}{\emph{Comput. Phys. Commun.}
  {\bfseries 183} (2012) 2158}
  [\href{https://arxiv.org/abs/1204.2314}{{\ttfamily 1204.2314}}].

\bibitem{Pak:2011xt}
A.~Pak, \emph{{The Toolbox of modern multi-loop calculations: novel analytic
  and semi-analytic techniques}},
  \href{https://doi.org/10.1088/1742-6596/368/1/012049}{\emph{J. Phys. Conf.
  Ser.} {\bfseries 368} (2012) 012049}
  [\href{https://arxiv.org/abs/1111.0868}{{\ttfamily 1111.0868}}].

\bibitem{Laporta:2001dd}
S.~Laporta, \emph{{High precision calculation of multiloop Feynman integrals by
  difference equations}}, \href{https://doi.org/10.1016/S0217-751X(00)00215-7,
  10.1142/S0217751X00002157}{\emph{Int. J. Mod. Phys.} {\bfseries A15} (2000)
  5087} [\href{https://arxiv.org/abs/hep-ph/0102033}{{\ttfamily
  hep-ph/0102033}}].

\bibitem{Lee:2012cn}
R.~N. Lee, \emph{{Presenting LiteRed: a tool for the Loop InTEgrals
  REDuction}},  \href{https://arxiv.org/abs/1212.2685}{{\ttfamily 1212.2685}}.

\bibitem{Smirnov:2014hma}
A.~V. Smirnov, \emph{{FIRE5: a C++ implementation of Feynman Integral
  REduction}}, \href{https://doi.org/10.1016/j.cpc.2014.11.024}{\emph{Comput.
  Phys. Commun.} {\bfseries 189} (2015) 182}
  [\href{https://arxiv.org/abs/1408.2372}{{\ttfamily 1408.2372}}].

\bibitem{Maierhoefer:2017hyi}
P.~Maierhöfer, J.~Usovitsch and P.~Uwer, \emph{{Kira—A Feynman integral
  reduction program}},
  \href{https://doi.org/10.1016/j.cpc.2018.04.012}{\emph{Comput. Phys. Commun.}
  {\bfseries 230} (2018) 99}
  [\href{https://arxiv.org/abs/1705.05610}{{\ttfamily 1705.05610}}].

\bibitem{Lee:2014ioa}
R.~N. Lee, \emph{{Reducing differential equations for multiloop master
  integrals}}, \href{https://doi.org/10.1007/JHEP04(2015)108}{\emph{JHEP}
  {\bfseries 04} (2015) 108} [\href{https://arxiv.org/abs/1411.0911}{{\ttfamily
  1411.0911}}].

\bibitem{Meyer:2016slj}
C.~Meyer, \emph{{Transforming differential equations of multi-loop Feynman
  integrals into canonical form}},
  \href{https://doi.org/10.1007/JHEP04(2017)006}{\emph{JHEP} {\bfseries 04}
  (2017) 006} [\href{https://arxiv.org/abs/1611.01087}{{\ttfamily
  1611.01087}}].

\bibitem{Meyer:2018feh}
C.~Meyer, \emph{{Algorithmic transformation of multi-loop Feynman integrals to
  a canonical basis}}, Ph.D. thesis, Humboldt U., Berlin, 2018-01-22.
\newblock \href{https://arxiv.org/abs/1802.02419}{{\ttfamily 1802.02419}}.
\newblock 10.18452/18763.

\bibitem{Gituliar:2017vzm}
O.~Gituliar and V.~Magerya, \emph{{Fuchsia: a tool for reducing differential
  equations for Feynman master integrals to epsilon form}},
  \href{https://doi.org/10.1016/j.cpc.2017.05.004}{\emph{Comput. Phys. Commun.}
  {\bfseries 219} (2017) 329}
  [\href{https://arxiv.org/abs/1701.04269}{{\ttfamily 1701.04269}}].

\bibitem{Prausa:2017ltv}
M.~Prausa, \emph{{epsilon: A tool to find a canonical basis of master
  integrals}}, \href{https://doi.org/10.1016/j.cpc.2017.05.026}{\emph{Comput.
  Phys. Commun.} {\bfseries 219} (2017) 361}
  [\href{https://arxiv.org/abs/1701.00725}{{\ttfamily 1701.00725}}].

\bibitem{Meyer:2017joq}
C.~Meyer, \emph{{Algorithmic transformation of multi-loop master integrals to a
  canonical basis with CANONICA}},
  \href{https://doi.org/10.1016/j.cpc.2017.09.014}{\emph{Comput. Phys. Commun.}
  {\bfseries 222} (2018) 295}
  [\href{https://arxiv.org/abs/1705.06252}{{\ttfamily 1705.06252}}].

\bibitem{Lee:2017oca}
R.~N. Lee and A.~A. Pomeransky, \emph{{Normalized Fuchsian form on Riemann
  sphere and differential equations for multiloop integrals}},
  \href{https://arxiv.org/abs/1707.07856}{{\ttfamily 1707.07856}}.

\bibitem{Remiddi:1999ew} 
E.~Remiddi and J.~A.~M.~Vermaseren, \emph{Harmonic polylogarithms},
\href{https://doi.org/10.1142/S0217751X00000367}{\emph{Int. J. Mod. Phys. A}
  {\bfseries 15} (2018) 725}
  [\href{https://arxiv.org/abs/hep-ph/9905237}{{\ttfamily hep-ph/9905237}}].


\bibitem{Maitre:2005uu}
D.~Maitre, \emph{{HPL, a mathematica implementation of the harmonic
  polylogarithms}},
  \href{https://doi.org/10.1016/j.cpc.2005.10.008}{\emph{Comput. Phys. Commun.}
  {\bfseries 174} (2006) 222}
  [\href{https://arxiv.org/abs/hep-ph/0507152}{{\ttfamily hep-ph/0507152}}].

\bibitem{Gehrmann-DeRidder:2003pne}
A.~Gehrmann-De~Ridder, T.~Gehrmann and G.~Heinrich, \emph{{Four particle phase
  space integrals in massless QCD}},
  \href{https://doi.org/10.1016/j.nuclphysb.2004.01.023}{\emph{Nucl. Phys.}
  {\bfseries B682} (2004) 265}
  [\href{https://arxiv.org/abs/hep-ph/0311276}{{\ttfamily hep-ph/0311276}}].

\bibitem{Panzer:2014caa}
E.~Panzer, \emph{{Algorithms for the symbolic integration of hyperlogarithms
  with applications to Feynman integrals}},
  \href{https://doi.org/10.1016/j.cpc.2014.10.019}{\emph{Comput. Phys. Commun.}
  {\bfseries 188} (2015) 148}
  [\href{https://arxiv.org/abs/1403.3385}{{\ttfamily 1403.3385}}].

\bibitem{Konishi:1978yx}
K.~Konishi, A.~Ukawa and G.~Veneziano, \emph{{A Simple Algorithm for QCD
  Jets}}, \href{https://doi.org/10.1016/0370-2693(78)90015-1}{\emph{Phys.
  Lett.} {\bfseries 78B} (1978) 243}.

\bibitem{Richards:1982te}
D.~G. Richards, W.~J. Stirling and S.~D. Ellis, \emph{{Second Order Corrections
  to the Energy-energy Correlation Function in Quantum Chromodynamics}},
  \href{https://doi.org/10.1016/0370-2693(82)90275-1}{\emph{Phys. Lett.}
  {\bfseries 119B} (1982) 193}.

\bibitem{Gehrmann:2011aa}
T.~Gehrmann, M.~Jaquier, E.~W.~N. Glover and A.~Koukoutsakis, \emph{{Two-Loop
  QCD Corrections to the Helicity Amplitudes for $H \to$ 3 partons}},
  \href{https://doi.org/10.1007/JHEP02(2012)056}{\emph{JHEP} {\bfseries 02}
  (2012) 056} [\href{https://arxiv.org/abs/1112.3554}{{\ttfamily 1112.3554}}].

\bibitem{Jin:2018fak}
Q.~Jin and G.~Yang, \emph{{Analytic Two-Loop Higgs Amplitudes in Effective
  Field Theory and the Maximal Transcendentality Principle}},
  \href{https://doi.org/10.1103/PhysRevLett.121.101603}{\emph{Phys. Rev. Lett.}
  {\bfseries 121} (2018) 101603}
  [\href{https://arxiv.org/abs/1804.04653}{{\ttfamily 1804.04653}}].

\bibitem{Shtabovenko:2016whf}
V.~Shtabovenko, \emph{{FeynHelpers: Connecting FeynCalc to FIRE and
  Package-X}}, \href{https://doi.org/10.1016/j.cpc.2017.04.014}{\emph{Comput.
  Phys. Commun.} {\bfseries 218} (2017) 48}
  [\href{https://arxiv.org/abs/1611.06793}{{\ttfamily 1611.06793}}].

\bibitem{Patel:2015tea}
H.~H. Patel, \emph{{Package-X: A Mathematica package for the analytic
  calculation of one-loop integrals}},
  \href{https://doi.org/10.1016/j.cpc.2015.08.017}{\emph{Comput. Phys. Commun.}
  {\bfseries 197} (2015) 276}
  [\href{https://arxiv.org/abs/1503.01469}{{\ttfamily 1503.01469}}].

\bibitem{Patel:2016fam}
H.~H. Patel, \emph{{Package-X 2.0: A Mathematica package for the analytic
  calculation of one-loop integrals}},
  \href{https://doi.org/10.1016/j.cpc.2017.04.015}{\emph{Comput. Phys. Commun.}
  {\bfseries 218} (2017) 66}
  [\href{https://arxiv.org/abs/1612.00009}{{\ttfamily 1612.00009}}].

\bibitem{Kilgore:2013gba}
W.~B. Kilgore, \emph{{One-loop single-real-emission contributions to $pp\to H +
  X$ at next-to-next-to-next-to-leading order}},
  \href{https://doi.org/10.1103/PhysRevD.89.073008}{\emph{Phys. Rev.}
  {\bfseries D89} (2014) 073008}
  [\href{https://arxiv.org/abs/1312.1296}{{\ttfamily 1312.1296}}].

\bibitem{Chetyrkin:2000yt}
K.~G. Chetyrkin, J.~H. Kuhn and M.~Steinhauser, \emph{{RunDec: A Mathematica
  package for running and decoupling of the strong coupling and quark masses}},
  \href{https://doi.org/10.1016/S0010-4655(00)00155-7}{\emph{Comput. Phys.
  Commun.} {\bfseries 133} (2000) 43}
  [\href{https://arxiv.org/abs/hep-ph/0004189}{{\ttfamily hep-ph/0004189}}].

\bibitem{Herren:2017osy}
F.~Herren and M.~Steinhauser, \emph{{Version 3 of RunDec and CRunDec}},
  \href{https://doi.org/10.1016/j.cpc.2017.11.014}{\emph{Comput. Phys. Commun.}
  {\bfseries 224} (2018) 333}
  [\href{https://arxiv.org/abs/1703.03751}{{\ttfamily 1703.03751}}].

\bibitem{Collins:1981uk}
J.~C. Collins and D.~E. Soper, \emph{{Back-To-Back Jets in QCD}},
  \href{https://doi.org/10.1016/0550-3213(81)90339-4}{\emph{Nucl. Phys.}
  {\bfseries B193} (1981) 381}.

\bibitem{Dokshitzer:1999sh}
Y.~L. Dokshitzer, G.~Marchesini and B.~R. Webber, \emph{{Nonperturbative
  effects in the energy energy correlation}},
  \href{https://doi.org/10.1088/1126-6708/1999/07/012}{\emph{JHEP} {\bfseries
  07} (1999) 012} [\href{https://arxiv.org/abs/hep-ph/9905339}{{\ttfamily
  hep-ph/9905339}}].

\bibitem{Gao:2019ojf}
A.~Gao, H.~T. Li, I.~Moult and H.~X. Zhu, \emph{{The Transverse Energy-Energy
  Correlator in the Back-to-Back Limit}},
  \href{https://arxiv.org/abs/1901.04497}{{\ttfamily 1901.04497}}.

\bibitem{Ebert:2018gsn}
M.~A. Ebert, I.~Moult, I.~W. Stewart, F.~J. Tackmann, G.~Vita and H.~X. Zhu,
  \emph{{Subleading Power Rapidity Divergences and Power Corrections for
  $q_T$}},  \href{https://arxiv.org/abs/1812.08189}{{\ttfamily 1812.08189}}.

\bibitem{Sjostrand:2014zea}
T.~Sjöstrand, S.~Ask, J.~R. Christiansen, R.~Corke, N.~Desai, P.~Ilten et~al.,
  \emph{{An Introduction to PYTHIA 8.2}},
  \href{https://doi.org/10.1016/j.cpc.2015.01.024}{\emph{Comput. Phys. Commun.}
  {\bfseries 191} (2015) 159}
  [\href{https://arxiv.org/abs/1410.3012}{{\ttfamily 1410.3012}}].

\bibitem{An:2018dwb}
F.~An et~al., \emph{{Precision Higgs Physics at CEPC}},
  \href{https://arxiv.org/abs/1810.09037}{{\ttfamily 1810.09037}}.

\bibitem{Abreu:1996na}
{\scshape DELPHI} collaboration, \emph{{Tuning and test of fragmentation models
  based on identified particles and precision event shape data}},
  \href{https://doi.org/10.1007/s002880050295}{\emph{Z. Phys.} {\bfseries C73}
  (1996) 11}.


\bibitem{Brun:1997pa}
R.~Brun and F.~Rademakers, \emph{{ROOT: An object oriented data analysis
  framework}}, \href{https://doi.org/10.1016/S0168-9002(97)00048-X}{\emph{Nucl.
  Instrum. Meth.} {\bfseries A389} (1997) 81}.
  
\bibitem{Adachi:1989ej}
{\scshape TOPAZ} collaboration, \emph{{Measurements of $\alpha^- s$ in $e^+ e^-$ Annihilation
at $\sqrt{s}=53.3$\,{GeV} and 59.5\,{GeV}}}, \href{https://doi.org/10.1016/0370-2693(89)90969-6}{\emph{Phys.
  Lett.} {\bfseries B227} (1989) 495}.
  
\bibitem{Skands:2014pea}
P.~Skands, S.~Carrazza and J.~Rojo, \emph{{Tuning PYTHIA 8.1: the Monash 2013
  Tune}}, \href{https://doi.org/10.1140/epjc/s10052-014-3024-y}{\emph{Eur.
  Phys. J.} {\bfseries C74} (2014) 3024}
  [\href{https://arxiv.org/abs/1404.5630}{{\ttfamily 1404.5630}}].

\bibitem{Catani:1990rr}
S.~Catani, B.~R. Webber and G.~Marchesini, \emph{{QCD coherent branching and
  semiinclusive processes at large x}},
  \href{https://doi.org/10.1016/0550-3213(91)90390-J}{\emph{Nucl. Phys.}
  {\bfseries B349} (1991) 635}.

\bibitem{Abdallah:2003xz}
{\scshape DELPHI} collaboration, \emph{{A Study of the energy evolution of
  event shape distributions and their means with the DELPHI detector at LEP}},
  \href{https://doi.org/10.1140/epjc/s2003-01198-0}{\emph{Eur. Phys. J.}
  {\bfseries C29} (2003) 285}
  [\href{https://arxiv.org/abs/hep-ex/0307048}{{\ttfamily hep-ex/0307048}}].

\end{thebibliography}

\providecommand{\href}[2]{#2}\begingroup\raggedright\endgroup

\end{document}